\documentclass[12pt]{article}
\usepackage{amsmath,amssymb,amsfonts}
\usepackage[dvips]{graphicx}
\usepackage{epsfig}

\usepackage[vcentermath]{youngtab}

\makeatletter \@addtoreset{equation}{section} \makeatother

\renewcommand{\baselinestretch}{1.2}

\newcommand{\be}{\begin{eqnarray}}
\newcommand{\ee}{\end{eqnarray}}

\newcommand{\bn}{\begin{enumerate}}
\newcommand{\en}{\end{enumerate}}

\begin{document}

\renewcommand{\thefootnote}{\alph{footnote}}

\begin{titlepage}

\begin{center}
\hfill {\tt SNUTP14-005}\\
\hfill {\tt KIAS-P14031}\\

\vspace{2cm}

{\Large\bf Super-Yang-Mills theories on $S^4\times\mathbb{R}$}

\vspace{2cm}

\renewcommand{\thefootnote}{\alph{footnote}}

{\large Jungmin Kim$^1$, Seok Kim$^1$, Kimyeong Lee$^2$ and Jaemo Park$^3$}

\vspace{1cm}

\textit{$^1$Department of Physics and Astronomy \& Center for
Theoretical Physics,\\
Seoul National University, Seoul 151-747, Korea.}\\

\vspace{0.2cm}

\textit{$^2$School of Physics, Korea Institute for Advanced Study,
Seoul 130-722, Korea.}\\

\vspace{0.2cm}

\textit{$^3$Department of Physics \& Postech Center for Theoretical Physics (PCTP),\\
Postech, Pohang 790-784, Korea.}\\

\vspace{1cm}

E-mails: {\tt kjmint82@gmail.com, skim@phya.snu.ac.kr, klee@kias.re.kr, jaemo@postech.ac.kr}

\end{center}

\vspace{2cm}

\begin{abstract}

We construct super-Yang-Mills theories on $S^4\times\mathbb{R}$, $S^4\times S^1$ and
$S^4\times$ interval with the field content of maximal SYM, coupled to boundary
degrees in the last case. These theories provide building blocks of the `5d uplifts' of
gauge theories on $S^4$, obtained by compactifying the 6d $(2,0)$ theory.
We pay special attention to the $\mathcal{N}=2^\ast$ theory on $S^4$.
We also explain how to construct maximal SYM on $S^5\times\mathbb{R}$, and
clarify when SYM theories can be put on $S^n\times\mathbb{R}$.

\end{abstract}

\end{titlepage}

\renewcommand{\thefootnote}{\arabic{footnote}}

\setcounter{footnote}{0}

\renewcommand{\baselinestretch}{1}

\tableofcontents

\renewcommand{\baselinestretch}{1.2}

\section{Introduction and summary}

Studying gauge theories on curved manifolds provides useful insights on their dynamics.
In particular, supersymmetric gauge theories on curved manifolds have been extensively
studied in recent years with various exact results. Important examples are Euclidean
super-Yang-Mills theories on spheres \cite{Blau:2000xg}. Some recently studied ones
are SYM on $S^2$ \cite{Benini:2012ui,Doroud:2012xw}, $S^3$ \cite{Jafferis:2010un,Hama:2010av},
$S^4$ \cite{Pestun:2007rz}, and $S^5$ \cite{Hosomichi:2012ek}. In this paper, we
study SYM on $S^n\times\mathbb{R}$, $S^n\times S^1$, or $S^n\times I$ (interval),
with a focus on the case with $n=4$.

Yang-Mills theories on $S^n\times\mathbb{R}$ are
relatively simple models in many ways. For instance, studies on the phases of Yang-Mills
theories on $S^3\times\mathbb{R}$ \cite{Aharony:2003sx} led to deep understandings on their
dynamics, and also on the AdS$_5$ gravity duals when they exist. On very general grounds,
$S^n\times\mathbb{R}$ is one of the simplest Lorentzian curved spaces to put the field theory on.
Supersymmetric gauge theories on $S^n\times S^1$ are also studied in great details.
Their partition functions are indices which count BPS states, often related to the
`superconformal indices' which count local BPS operators of SCFTs \cite{Romelsberger:2005eg,Kinney:2005ej,Bhattacharya:2008zy}. There
have been extensive studies on these indices in various dimensions: for instance,
on $S^2\times S^1$ \cite{Kim:2009wb,Imamura:2011su}, $S^3\times S^1$
\cite{Kinney:2005ej,Romelsberger:2005eg}, $S^4\times S^1$ \cite{Kim:2012gu,Iqbal:2012xm},
$S^5\times S^1$
\cite{Kim:2012ava,Kallen:2012zn,Kim:2012tr,Lockhart:2012vp,Kim:2012qf,Kim:2013nva}.\footnote{
The index on $S^1\times S^1$, or a 2-torus, has a longer history.
This index is called the elliptic genus. For SUSY gauge theories, the elliptic genera were
studied rather recently in \cite{Gadde:2013dda,Benini:2013nda}.} Super-Yang-Mills theories on
$S^n\times S^1$ (or sometimes on different manifolds) related to the SCFT's
are often used to compute them.

Apart from the case with $n=3$, classical Yang-Mills theory carries an intrinsic scale,
the coupling constant $g_{YM}$. So there is no canonical way of writing down its action
on $S^n\times\mathbb{R}$, although the manifold is conformally flat. Demanding
certain SUSY provides strong constraints on possible SYM action on $S^n\times\mathbb{R}$.
However, a systematic study on writing down these SYM action appears unexplored in
some dimensions, at least not as much as the SYM on $S^n$. In fact,
the relatively well-known SYM theories on $S^n$ provide strong constraints on the possible
SYM theories on $S^n\times S^1$ via the small $S^1$ limit. This also constrains the SYM on
$S^n\times\mathbb{R}$, and the bulk term of the SYM on $S^n\times I$. We would like to
clarify this issue in various dimensions.

In particular, we focus on the SYM on $S^4\times\mathbb{R}$, $S^4\times S^1$,
and $S^4\times I$ in this paper. One motivation is that the 5 dimensional (maximal)
SYM theory is useful to study the dynamics of 6d $(2,0)$ superconformal
field theory \cite{Witten:1995zh,Strominger:1995ac,Witten:1995em} with circle
compactification, often by studying the non-perturbative sector
of the 5d SYM \cite{Douglas:2010iu,Lambert:2010iw,Bern:2012di,Papageorgakis:2014dma}.
Nonperturbative studies of SYM on $S^4\times S^1$ or $S^4\times\mathbb{R}$ could thus
shed light on the 6d $(2,0)$ theory on $S^4\times T^2$ or $S^4\times$ cylinder, just
like the similar studies on $\mathbb{R}^4\times S^1$ allowed one to study 6d
theory on $\mathbb{R}^4\times T^2$ \cite{Kim:2011mv,Haghighat:2013gba}.
\cite{Alday:2009aq,Wyllard:2009hg} considered the $SU(N)$
$(2,0)$ theory on $S^4\times\Sigma_2$, where $\Sigma_2$ is a Riemann surface,
with some punctures (codimension $2$ defects). They found that the gauge theory
partition functions on $S^4$ map to observables of the Liouville/Toda CFTs on $\Sigma_2$.
The 5d SYM on $S^4\times\mathbb{R}$ may provide some insights on this relation.
From the viewpoint of 5d SYM, the KK modes of the Liouville/Toda theories
on a cylinder should be visible as the nonperturbative instantonic particles on $S^4$. Even
without instantons, it would be interesting to see if reducing SYM theory on a small $S^4$
yields the Liouville/Toda quantum mechanics.

With these questions in mind,
we focus on a more elementary problem, to clearly show that it is possible to put the
$(2,0)$ theory on $S^4\times\mathbb{R}^2$ preserving some SUSY. After compactifying one of
the two directions of $\mathbb{R}^2$ to a circle, maximal SYM on $S^4\times\mathbb{R}$
should also exist, preserving some SUSY. This SYM on $S^4\times\mathbb{R}$ has not been
constructed yet, which we do in this paper. Also, general (punctured) Riemann surface has
limits in its moduli space.
The surface consists of long `tubular regions,' whose boundaries are either connected by
the 3-point junctions or end on the punctures. The limit corresponds to a weak coupling
limit of the 4d theory \cite{Gaiotto:2009we}. In this paper, we also
construct the 5d SYM living on the tubular region, namely on $S^4\times I$ (interval) after
circle reduction. We also find its coupling with boundary degrees living on $S^4$.

Let us explain the basic idea of constructing the SYM theory on $S^4\times\mathbb{R}$,
after which one can also replace $\mathbb{R}$ by $S^1$ or $I$. Perhaps we can start by
providing a resolution of a puzzle phrased in \cite{Terashima:2012ra}, which also arises
for SYM on general $S^n\times\mathbb{R}$. \cite{Terashima:2012ra} attempted to
construct 5d $\mathcal{N}=1$ SYM on $S^4\times S^1$ with a vector supermultiplet,
and reported a failure. One way to understand this failure is as follows. The 4d vector
multiplet of the $\mathcal{N}=2$ SYM on $S^4$ with radius $r$ contains two real scalars,
which have nonzero mass-square $\frac{2}{r^2}$.
Trying to find a 5d uplift of it on $S^4\times S^1$, one of
the two 4d scalars should uplift to the $S^1$ component $A_5$ of the gauge field, which
should have zero 5d mass from gauge symmetry. As $A_5$ transforms trivially under
all the global symmetries, it is impossible to induce a nonzero 4d mass
to $A_5$ via Scherk-Schwarz-like compactification. This appears to make it impossible to
realize minimal SYM on $S^4\times S^1$ which reduces to pure $\mathcal{N}=2$ SYM on
$S^4$. It also appears that 5d $\mathcal{N}=1$ SYM coupled to hypermultiplets in general
representation of the gauge group cannot exist, for the same reason.

We find a SYM on $S^4\times\mathbb{R}$ when the field content is the maximal vector
supermultiplet, consisting of 5d $\mathcal{N}=1$ vector multiplet and an adjoint hypermultiplet.
This theory preserves $8$ real SUSY. Reducing it on a small circle, we obtain a special
$\mathcal{N}=2^\ast$ theory on $S^4$ of \cite{Pestun:2007rz}, in which the hypermultiplet
mass parameter is specially tuned. The tuning is such that the curvature-coupling mass
contribution is balanced with the extra $\mathcal{N}=2^\ast$ mass contribution, yielding
zero net mass for two scalars in the 4d hypermultiplet. One of these two massless 4d scalars
uplifts to the $A_5$ component of the 5d gauge field, and another remains to be a
massless scalar in 5d. So the puzzle phrased in \cite{Terashima:2012ra} is resolved by providing
the massless $A_5$ from a 4d hypermultiplet scalar. Of course one should be able to realize
general $\mathcal{N}=2^\ast$ mass on $S^4$ by a reduction from the 5d/6d system. Or more
generally, one would like to find a higher dimensional uplift of the 4d SYM theories on $S^4$
with the field contents of \cite{Gaiotto:2009we}. (At least this is naturally suggested by the AGT
correspondence.) We find that the general $\mathcal{N}=2^\ast$ theory of \cite{Pestun:2007rz} can
be uplifted to the SYM on $S^4\times S^1$ with a defect wrapping $S^4$ and localized on $S^1$.
This defect uplifts in 6d to a puncture on the Riemann surface ($T^2$ in this case), which is
natural from the construction of \cite{Gaiotto:2009we}. Some theories on $S^4$ with
field contents discussed in \cite{Gaiotto:2009we} can be `uplifted
to 5d' by taking many SYM on $S^4\times I$, connecting various intervals and coupling the 5d
theories to various 4d degrees at the boundaries of $I$. The construction is well motivated by the D4-NS5 systems of \cite{Witten:1997sc}.

As the setup of AGT is wrapping the 6d $(2,0)$ theory on $S^4$, it only demands the
existence of a SYM on $S^4\times \mathbb{R}$ with the field content of maximal SYM.
We have no ideas on other 5d SYM on $S^4\times\mathbb{R}$.

One could in principle obtain a quantum mechanical description of this system
when $S^4$ is small. AGT correspondence could be suggesting that we shall obtain
the Liouville/Toda quantum mechanics. We only make a few comments on it in section 3.
It appears that non-perturbative effects of the 5d SYM should play important
roles to fully visualize the Liouville physics, even in the quantum mechanical version.

Although the main focus of this paper is the SYM theories on $S^4\times\mathbb{R}$,
we overview the problem of constructing supersymmetric Yang-Mills theory on
$S^n\times\mathbb{R}$ in various dimensions, also summarizing known results.
Just like the case of $S^4\times\mathbb{R}$, a constraint emerges from the scalar masses
on $S^n$ after compactifying $\mathbb{R}$ to a small
$S^1$. We summarize known SYM theories on various $S^n$ and $S^n\times\mathbb{R}$, and
also find new maximal SYM on $S^5\times\mathbb{R}$.
The SYM on $S^n\times\mathbb{R}$ with $n\geq 6$ appears to be forbidden. We also
discuss possible applications of these theories.

The rest of this paper is organized as follows. In section 2, we construct the SYM
on $S^4\times\mathbb{R}$, $S^4\times S^1$ and $S^4\times I$ with boundary degrees.
In section 3, we make a few remarks on the mechanical system obtained by taking $S^4$
to be small. In section 4, we consider the possibilities of SYM theories on
various $S^n\times\mathbb{R}$, explain that maximal SYM exist for $n=5$, and
comment on its possible applications.

\section{SYM on $S^4\times\mathbb{R}$}

We start by providing a simple argument for the existence of a SYM on $S^4\times S^1$ with a
maximal vector supermultiplet. This can be easily seen by starting from a 4d deconstruction
description of the 6d $U(N)$ $(2,0)$ theory on $T^2$ \cite{ArkaniHamed:2001ie}.\footnote{This
description works for $U(N)$ gauge group. We take the arguments below as a
guidance for $U(N)$, while the actual construction of 5d SYM on $S^4\times S^1$ is made with
arbitrary gauge group.} The deconstructed theory is given by a 4d $\mathcal{N}=2$ superconformal
field theory, described by a circular quiver diagram of $U(N)^K$ vector multiplet and
bi-fundamental hypermultiplets for adjacent $U(N)$ pairs in the quiver. One starts from this 4d
theory and give nonzero VEV to the $K$ hypermultiplets, which spontaneously breaks $U(N)^K$ to
$U(N)$. This Higgsing triggers an RG flow, and taking a suitable large $K$ scaling limit is
suggested to yield the 6d $(2,0)$ theory on $T^2$.

The 4d classical gauge theory is obtained by deconstructing classical 5d maximal SYM on $S^1$
\cite{Lambert:2012qy}. Discretizing the circle direction, one obtains the
expected $U(N)^K$ circular quiver theory in its Higgs branch. Thus, the large $K$ limit of the $\mathcal{N}=2$ superconformal theory on $\mathbb{R}^4$ yields classical
maximal SYM on $\mathbb{R}^4\times S^1$. The 4d fields which acquire nonzero masses via
Higgs mechanism provide the infinite tower of Kaluza-Klein modes on $S^1$ in the
large $K$ limit. The discussions of \cite{Lambert:2012qy} are mostly within the classical
field theory, so that it can be applied to maximal SYM on any $\mathbb{R}^n\times S^1$,
supposing that $\mathbb{R}^n$ admits SYM with $8$ SUSY. Namely, after discretizing the fields
along $S^1$ as \cite{Lambert:2012qy}, one would obtain an $n$ dimensional SYM with $8$ SUSY
described by a $U(N)^K$ circular quiver. We focus on the case with $n=4$ here, commenting on
other dimensions in section 4.

The above procedure on $\mathbb{R}^4\times S^1$ can be generalized to SYM on $S^4\times S^1$.
Firstly, note that the above 4d superconformal quiver theory can be put on $S^4$ with radius $r$,
as the latter space is conformally flat. All the scalars in the hypermultiplet acquire
conformal mass-square $\frac{2}{r^2}$. So at this stage, one cannot Higgs this theory, and
thus cannot address the 6d $(2,0)$ on $S^4\times T^2$ or 5d SYM on $S^4\times S^1$.
What we need is a mass-deformation of the CFT on $S^4$, with an extra mass
parameter for the 4d $\mathcal{N}=2$ hypermultiplets. This deformation is basically the same as
that in \cite{Pestun:2007rz} for the $\mathcal{N}=2^\ast$ theory on $S^4$, and for general
field contents can be derived from \cite{Hama:2012bg}.
The mass parameter can be tuned to have two of the four scalars in a hypermultiplet
to be massless, as we shall explain below shortly. We set the mass parameter to this
value. Now the $K$ Higgs fields can acquire expectation values, by turning on one of the two
massless scalars per hypermultiplet. Then we have exactly the same mechanism as \cite{Lambert:2012qy},
obtaining the Kaluza-Klein modes for the 5d SYM on $S^4\times S^1$ in the large $K$ limit.
Another massless scalar is identified as the 5d gauge field $A_5$ along the circle. The last
identification is possible as this scalar always appears in the 4d action with derivatives
or in commutators, because this scalar plays the role of `would-be Goldstone boson' for
the broken $U(N)^{K-1}$ gauge symmetry.

The details of the 5d theory can also be obtained by deconstruction methods, although
it could be a bit cumbersome. We find the above existence argument itself quite useful.
We shall construct this theory in the next subsection more efficiently with arbitrary
gauge group, using the off-shell supergravity method of \cite{Cordova:2013bea}.

The theory constructed this way on $S^4\times S^1$ has its 4d reduction
given by a special $\mathcal{N}=2^\ast$ theory on $S^4$, with the adjoint hypermultiplet
mass parameter tuned to have two massless scalars. To compare
with the 5d theory we construct later, let us consider this special $\mathcal{N}=2^\ast$
theory on $S^4$. The general mass-square matrix for the hypermultiplet
scalar is \cite{Pestun:2007rz}\footnote{The coefficient of the last term was $-\frac{1}{4r}$
in \cite{Pestun:2007rz}, instead of $-\frac{1}{r}$ that we wrote. We find that $-\frac{1}{r}$
is correct, by following the derivations of \cite{Pestun:2007rz}. Namely, we find
$$
  \frac{1}{2r}\Psi\Gamma^i\Gamma^{kl}\varepsilon R_{kl}M_{ij}\Phi_j=
  \frac{2}{r}(\Psi\Gamma^i\varepsilon)R_{ki}M_{kj}\Phi_j
$$
at the second step of eqn.(2.23) of \cite{Pestun:2007rz}, where the right hand side
is $4$ times what is written in \cite{Pestun:2007rz}.}
\begin{equation}
  \frac{2}{r^2}\delta_{ij}-M_{ik}M_{jk}-\frac{1}{r}R_{k(i|}M_{k|j)}\ .
\end{equation}
Here $i,j=5,6,7,8$ label four real scalars, $M_{ij}$ is an $SU(2)^R_R$ rotation matrix in
$SO(4)\subset SO(6)_R$, and $R_{ij}$ is an $SU(2)^R_L$ (i.e. anti-self-dual) element normalized
as $R^{kl}R^{kl}=4$ \cite{Pestun:2007rz}. We can take
\begin{equation}\label{pestun-matrices}
  R=\left(\begin{array}{cccc}&&0&1\\&&-1&0\\0&1&&\\
  -1&0&&\end{array}\right)\ ,\ \ M=m\left(\begin{array}{cccc}&&0&1\\&&1&0\\
  0&-1&&\\-1&0&&\end{array}\right)\ .
\end{equation}
The convention for $m$ is same as that used in section 4 of \cite{Pestun:2007rz},
in which $M_{ij}M^{ij}=4m^2$. The mass-square eigenvalues are
\begin{equation}
  \frac{2}{r^2}-m^2\pm\frac{m}{r}
\end{equation}
where an eigenvalue with given sign appears twice in the matrix.
At the point $m=0$ with maximal SUSY, all four scalars of the hypermultiplet have the
conformal mass-square $\frac{2}{r^2}$ (same as the two scalars in the 4d vector multiplet).
On the other hand, at $m=\pm\frac{1}{r}$, two of the four scalars have conformal
mass-square $\frac{2}{r^2}$, while the other two are massless. This mass matrix with
$m=\pm \frac{1}{r}$ is what we shall find from the circle reduction of our 5d SYM
on $S^4\times S^1$, with one of the massless scalars uplifting to
$A_5$ component of the 5d vector potential on $S^1$.

Similar analysis can be done for the $U(N)^K$ circular quiver gauge theory,
by using the results of \cite{Hama:2012bg}. This guarantees that one can Higgs
the theory at $m=\pm\frac{1}{r}$ and deconstruct the 5d SYM on $S^4\times S^1$. We
do not elaborate on it here.

We also explain the Killing spinor equation of the $\mathcal{N}=2^\ast$ theory on
$S^4$ \cite{Pestun:2007rz}, which will be compared to what we shall obtain from our
5d SYM on $S^4\times S^1$. The spinors in \cite{Pestun:2007rz} are written in 10d
$\mathcal{N}=1$ notation, while we shall naturally use its 5d reduction, which is a spinor in
Lorentz group $SO(5)$ and R-symmetry group $SO(5)_R$.\footnote{In \cite{Pestun:2007rz}, $SO(9,1)$ spinors were used, with $({\bf \Gamma}^0)^2=-1$ for an internal direction. Having in mind the
continuation with Euclidean R-symmetry, we put an extra $i$ factor to ${\bf \Gamma}^0$. However,
whenever we discuss Majorana spinors in 10d, this will essentially be in the Minkowskian sense
as in \cite{Pestun:2007rz}, $\bar{\bf \Psi}={\bf \Psi}^TC_{10}$. See appendix.} We find it
convenient to introduce the following $32\times 32$ gamma matrices ${\bf \Gamma}^M$ in 10d,
using our $4\times 4$ ones $\Gamma^\mu$ (for 5d space), and $\hat\Gamma^I$ (for $SO(5)_R$):
\begin{eqnarray}\label{5d-10d-gamma}
  {\bf \Gamma}^\mu&=&\Gamma^\mu\otimes\hat\Gamma^5\otimes\sigma_1\ \ ,\ \ \
  {\bf \Gamma}^{i+5}={\bf 1}_4\otimes i\hat\Gamma^{5i}\otimes\sigma_1\ \ \ (i=1,2,3)\nonumber\\
  {\bf \Gamma}^9&=&{\bf 1}_4\otimes i\hat\Gamma^{54}\otimes\sigma_1\ \ ,\ \ \
  {\bf \Gamma}^0={\bf 1}_4\otimes{\bf 1}_4\otimes\sigma_2\ .
\end{eqnarray}
We also define the 10d chirality  operator
${\bf \Gamma}^{11}=-i{\bf \Gamma}^{1234567890}={\bf 1}_4\otimes{\bf 1}_4\otimes\sigma_3$.
We shall be working with 5d gamma matrices satisfying $\Gamma^{12345}=1$, $\hat\Gamma^{12345}=1$.
The 10d $\mathcal{N}=1$ SUSY satisfies $\Gamma^{11}\epsilon=\epsilon$, or
\begin{equation}
  \sigma_3\epsilon=\epsilon\ .
\end{equation}
Furthermore, the 8 supercharges of the 4d $\mathcal{N}=2^\ast$
theory satisfy the projection \cite{Pestun:2007rz}
\begin{equation}
  {\bf \Gamma}^{5678}\epsilon=\epsilon\ ,
\end{equation}
where the $5678$ directions are for the four scalars in the adjoint hypermultiplet
from the viewpoint of 4d SYM. $9$ and $0$ directions are for the two real scalars in
the 4d vector multiplet. From our 5d SYM on $S^4\times S^1$, ${\bf \Gamma}^5$
is for the fifth spatial direction which we take to be $S^1$, and the remaining $678$
are for the first three of the five internal directions. In particular, we find that
\begin{equation}\label{N=2*-projection}
   \epsilon={\bf \Gamma}^{5678}\epsilon
   =i\Gamma^5\hat\Gamma^{123}\epsilon=-i\Gamma^5\hat\Gamma^{45}\epsilon\ .
\end{equation}
The Killing spinor equation on $S^4$, in the $(10,0)$ signature, is given by \cite{Pestun:2007rz}
\begin{equation}
  \nabla_a\epsilon=-\frac{i}{8r}{\bf \Gamma}_a{\bf \Gamma}^{0kl}R_{kl}\epsilon\ ,
\end{equation}
where indices run over $k,l=5,6,7,8$, $R$ can be chosen as (\ref{pestun-matrices}), and
$a=1,2,3,4$. This equation has $8$ solutions, which generate $OSp(2|4)$ supersymmetry.
Using (\ref{N=2*-projection}), one obtains
\begin{equation}\label{4d-projection}
  {\bf \Gamma}_\mu{\bf \Gamma}^{0kl}R_{kl}\epsilon=
  2{\bf\Gamma}_\mu{\bf \Gamma}^0({\bf \Gamma}^{58}-{\bf \Gamma}^{67})\epsilon
  =4i\Gamma_\mu\hat\Gamma^{34}\epsilon\ .
\end{equation}
Thus in our 5d notation, the $S^4$ Killing spinor equation is given by
\begin{equation}\label{4d-killing}
  \nabla_\mu\epsilon=\frac{1}{2r}\Gamma_\mu\hat\Gamma^{34}\epsilon\ \ \
  (\mu=1,2,3,4)\ .
\end{equation}
This is what we shall obtain from the SYM on $S^4\times\mathbb{R}$, together with
$\partial_5\epsilon=0$.

\subsection{Construction from off-shell supergravity}

We construct the maximal SYM on $S^4\times\mathbb{R}$ using supergravity methods of
\cite{Cordova:2013bea}. Although it is straightforward to uplift the 4d SYM to $S^4\times S^1$
with a massless scalar, there are benefits for constructing it using
the formalism of \cite{Cordova:2013bea}. The most important point is that our construction
below will not be just finding 5d SYM on $S^4\times\mathbb{R}$, but will also specify
the $S^4\times\mathbb{R}^2$ supergravity background on which one can put the $(2,0)$ theory.
One may be interested in studying a 5d SYM obtained by a different circle reduction.

We first construct the off-shell supergravity background $S^4\times S^1$ or $S^4\times\mathbb{R}$, admitting Killing spinors, and then write down an on-shell SUSY action in that background. The SUSY condition for the gravitino requires
\begin{equation}
  D_\mu\epsilon^m=\frac{i}{2}S^{mn}\Gamma_\mu\epsilon_n=-\frac{i}{2}S^m_{\ \ n}\Gamma_\mu\epsilon^n\ ,
\end{equation}
with $\mu=1,2,3,4,5$, where
$D_\mu\epsilon^m=\nabla_\mu\epsilon^m-\frac{1}{2}(V_\mu)^m_{\ n}\epsilon^n$.
Here, $V_\mu$ is the background gauge field for the $SO(5)_R$ symmetry. $S$
is an $SO(5)_R$ adjoint, or $Sp(4)$ antisymmetric, scalar which comes from
the circle reduction of the $SO(5)_R$ gauge field in 6d. $m,n=1,2,3,4$ are
$SO(5)_R$ spinor indices. See \cite{Cordova:2013bea} for more on notations. We also write
\begin{equation}
  S^{mn}=S^{IJ}(\hat\Gamma^{IJ})^{mn}\ ,\ \ V_\mu^{mn}=V_\mu^{IJ}(\hat\Gamma^{IJ})^{mn}
\end{equation}
with $I,J=1,\cdots,5$ being the $SO(5)_R$ vector indices.
In foresight, let us turn on nonzero
$S^{34}$ and $V_5^{35}$ in the last $IJ$ basis. This setting will turn out to admit backgrounds
which preserve $8$ real SUSY, both on $S^4\times S^1$ and $S^4\times\mathbb{R}$. The above Killing
spinor equation becomes
\begin{eqnarray}\label{5d-killing}
  \nabla_a\epsilon&=&-iS^{34}\hat\Gamma^{34}\Gamma_a\epsilon\ ,\nonumber\\
  \left(\partial_5-V^{35}_5\hat\Gamma^{35}\right)\epsilon&=&
  -iS^{34}\hat\Gamma^{34}\Gamma_5\epsilon\ ,
\end{eqnarray}
with $a=1,2,3,4$. Integrability on the $S^4$ part demands
\begin{equation}
  S^{34}=\pm\frac{i}{2r}\ .
\end{equation}
To be definite, let us choose $S^{34}=+\frac{i}{2r}$. So we obtained
a \textit{complexified} background for the scalar $S^{IJ}$. Then, demanding the spinor
to be constant on $S^1$ or $\mathbb{R}$, one obtains
\begin{equation}\label{projection}
  V^{35}_5=S^{34}=\frac{i}{2r}\ ,\ \ \hat\Gamma^{45}\Gamma_5\epsilon=i\epsilon\ .
\end{equation}
Again, we chose a definite sign between two possibilities. Most generally,
one obtains four possibilities, depending on the two signs of $V_5^{35}$ and $S^{34}$.
These will correspond to having two possible values $m=\pm\frac{1}{r}$ for the 4d
hypermultiplet mass after the circle reduction, and also the $\pm$ signs on the right
hand side of (\ref{4d-killing}). The projection condition (\ref{projection}) for
$\hat\Gamma^{45}\Gamma_5$ is consistent with the $S^4$ part of the equation, as both
$\nabla_a$ on the left hand side and $\hat\Gamma^{34}\Gamma_a$ commute with
$\hat\Gamma^{45}\Gamma_5$. This projection reduces the spinor components of $\epsilon$
from $16$ to $8$. One may ask whether the remaining $8$ components with
$\hat\Gamma^{45}\Gamma_5\epsilon=-i\epsilon$ could solve the second Killing spinor equation
on $S^4\times\mathbb{R}$, depending on $x^5$. We find no such solutions which are compatible
with the first equation of (\ref{5d-killing}).
So this background preserves $8$ SUSY on both $S^4\times S^1$ and $S^4\times\mathbb{R}$. Note
that, the $S^4$ part of (\ref{5d-killing}) and the projection in (\ref{projection})
are the same as (\ref{4d-killing}), (\ref{N=2*-projection}) for the $\mathcal{N}=2^\ast$
theory on $S^4$.

Before proceeding, we turn to an issue of the reality condition on spinors.
In \cite{Cordova:2013bea}, all Lorentzian fermions are taken to satisfy symplectic-Majorana
conditions. The matter fermion and Killing spinor satisfy the same reality condition.
Let us discuss the reality condition for $\epsilon$ here. The reality condition
is $\bar\epsilon=\epsilon^TC\Omega$, where $C,\Omega$ satisfy
$C\Gamma_\mu^TC^{-1}=\Gamma_\mu$, $\Omega(\hat\Gamma^I)^T\Omega^{-1}=\hat\Gamma^I$. To be
concrete, we assume
\begin{equation}\label{SO(5)-gamma}
  \Gamma^\mu=\left(\begin{array}{cc}0&\sigma^a\\ \bar\sigma^a&0\end{array}\right)\ ,\ \
  \left(\begin{array}{cc}{\bf 1}_2&0\\0&-{\bf 1}_2\end{array}\right)
\end{equation}
with $a=1,2,3,4$, $\sigma^a=(-i\vec\tau,1)$, $\bar\sigma^a=(i\vec\tau,1)$, and
\begin{equation}\label{SO(5)R-gamma}
  \hat\Gamma^I=\left(\begin{array}{cc}0&\sigma^m\\ \bar\sigma^m&0\end{array}\right)\ ,\ \
  \left(\begin{array}{cc}{\bf 1}_2&0\\0&-{\bf 1}_2\end{array}\right)
\end{equation}
with $m=1,2,3,4$, $\sigma^m=(-i\vec\tau,1)$, $\bar\sigma^m=(i\vec\tau,1)$. Then we can
take $C=-\Gamma^{13}={\rm diag}(\epsilon,\epsilon)$ with $\epsilon^{12}=-\epsilon^{21}=1$,
and $\Omega=-\hat\Gamma^{13}={\rm diag}(\epsilon,\epsilon)$. Had it been the Lorentzian
theory in a real background for $V,S$, the reality condition would come
with $\bar\epsilon=\epsilon^\dag\Gamma^0$. In this case, the SUSY condition from the
gravitino variation $\delta\psi_\mu$ and its conjugate $\delta\bar\psi_\mu$ are equivalent
so that solving the former (\ref{5d-killing}) suffices. However, going to Euclidean
signature and having a complex background both affect the equivalence. For the consistency
of our analysis above, we should carefully choose the definition of $\bar\epsilon$ so that
solving (\ref{5d-killing}) still suffices in our Euclidean complex background.
Namely, starting from (\ref{5d-killing}), we derive the equations for $\epsilon^TC\Omega$
and $\bar\epsilon\equiv\epsilon^\dag M$, and require the two to be the same. This
imposes the following conditions on $M$:
\begin{equation}
  0=[M,\hat\Gamma^{34}\Gamma_a]=[M,\Gamma^{ab}]=[M,\hat\Gamma^{34}\Gamma_5]=
  \{M,\hat\Gamma^{35}\}\ .
\end{equation}
These conditions are satisfied by $M\propto\hat\Gamma^5$.
We take $\bar\epsilon\equiv\epsilon^\dag(-\hat\Gamma^5)$, and the same
definition for barred fermions holds for matters below.

To complete the construction of the SUSY background, we also consider the dilatino
equation of \cite{Cordova:2013bea} with nonzero $V^{35},S^{34},D^{mn}_{rs}$. This is
given, in Euclidean signature (in which we Wick-rotate from the Lorentzian theory with
$\epsilon^{01234}=1$ by $x^0=-ix^5$), by
\begin{equation}
  \delta\chi^{mn}_r=-\frac{i}{12}D^\lambda S_r^{\ [m}
  \varepsilon_{\mu\nu\rho\sigma\lambda}\Gamma^{\mu\nu\rho\sigma}\epsilon^{n]}
  -\frac{4}{15}D^{mn}_{rs}\epsilon^s-({\rm trace})=0\ ,
\end{equation}
where $D_\mu S=\partial_\mu S-\frac{1}{2}[V_\mu, S]$. The subtracted `trace' terms are
explained in \cite{Cordova:2013bea}, related to $D^{mn}_{rs}$ satisfying
$0=D^{mn}_{rs}\Omega_{mn}=D^{mn}_{rs}\Omega^{rs}=
D^{mn}_{mn}$. The solution to this equation is
\begin{equation}
  D^{mn}_{rs}=-\frac{15}{2r^2}\left[(\hat\Gamma^{45})_r^{[m}(\hat\Gamma^{45})^{n]}_s
  -\frac{1}{5}\delta^{[m}_r\delta^{n]}_s-\frac{1}{5}\Omega^{mn}\Omega_{rs}\right]\ ,
\end{equation}
where we have used our convention $\Gamma^{12345}=1$ for the gamma matrices.
(In all four cases in which $S^{34}$, $V^{35}_5$ take $\pm$ signs, the above solution
for $D^{mn}_{rs}$ is always the same.) This completes the construction of the 5d supergravity
background. We note that one can easily uplift this 5d background to the 6d supergravity
background on $S^4\times \mathbb{R}^2$, following \cite{Cordova:2013bea}.

Once the background is found, the SYM action on $S^4\times\mathbb{R}$ or
$S^4\times S^1$ immediately follows from the results of \cite{Cordova:2013bea}.
Our Euclidean theory is obtained by a Wick rotation from theirs, on the fields
and the $x^0$ coordinate. The action is given by
\begin{eqnarray}\label{action}
  S\!&\!=\!&\!\frac{1}{g_{YM}^2}\int d^5x\sqrt{g}\
  {\rm tr}\left[\frac{1}{4}F_{\mu\nu}F^{\mu\nu}\!
  +\frac{1}{2}D_\mu\varphi^ID^\mu\varphi^I\!-\frac{1}{4}[\varphi^I,\varphi^J]^2+
  \frac{1}{r^2}(\varphi^i)^2+\frac{1}{r^2}(\varphi^{i^\prime})^2\right.\\
  &&\hspace{2cm}\left.-\frac{2i}{r}\varphi^5\left(D_5\varphi^3-i[\varphi^1,\varphi^2]\right)
  +\frac{i}{2}\bar\Psi\Gamma^\mu D_\mu\Psi
  +\frac{i}{2}\bar\Psi\hat\Gamma^I[\phi^I,\Psi]-\frac{i}{4r}\bar\Psi
  \left(\hat\Gamma^{34}+i\hat\Gamma^{35}\Gamma^5\right)\Psi\right]\nonumber
\end{eqnarray}
where $I=1,2,3,4,5$, $i=4,5$, $i^\prime=1,2$ for the $SO(5)_R$ vector.
Again $\bar\Psi\equiv\Psi^\dag(-\hat\Gamma^5)$, and all $SO(5)_R$ spinor contractions
above are understood as $\bar\Psi_m(\cdots)\Psi^m$, $\bar\Psi_m(\hat\Gamma^I)^m_{\ \ n}\Psi^n$, etc. We also note that our derivatives $D_5$ are just gauge covariant
derivative of SYM, not covariantized with background $V_5$ gauge field for $SO(5)_R$.
The SUSY transformations are given by
\begin{eqnarray}
  \delta A_\mu&=&-i\bar\epsilon_m\Gamma_\mu\Psi^m\nonumber\\
  \delta\varphi^I&=&\bar\epsilon_m(\hat\Gamma^I)^m_{\ \ n}\Psi^n\\
  \delta\Psi^m&=&\frac{1}{2}F_{\mu\nu}\Gamma^{\mu\nu}\epsilon^m
  +i\Gamma^\mu D_\mu\varphi^I(\hat\Gamma^I)^m_{\ n}\epsilon^n\!
  +\!\frac{i}{r}\left(\phi^i(\hat\Gamma^i\hat\Gamma^{34})^m_{\ n}+2\phi^{i^\prime}
  (\hat\Gamma^{i^\prime}\hat\Gamma^{34})^m_{\ n}\right)\epsilon^n\!
  -\!\frac{i}{2}[\varphi^I,\varphi^J](\hat\Gamma^{IJ})^m_{\ n}\epsilon^n\nonumber\ .
\end{eqnarray}
$\Psi$ satisfies the same reality condition as $\epsilon$, $\bar\Psi=\Psi^TC\Omega$.

Since the $8$ SUSY satisfies the projection condition
$\hat\Gamma^{45}\Gamma_5\epsilon=i\epsilon$, one can decompose the fermion $\Psi$
into two parts: $\lambda$ which has $+i$ eigenvalue of this matrix, and $\psi$ which
has $-i$ eigenvalue. The SUSY transformation then naturally divides the 5d maximal
vector multiplet into 4d $\mathcal{N}=2$ `vector multiplet' $A_a,\lambda,\varphi^{4,5}$
(with $a=1,2,3,4$) and `hypermultiplet' $A_5,\varphi^{1,2,3},\psi$. The SUSY
transformation rules are
\begin{eqnarray}
  \delta A_a&=&-i\bar\epsilon_m\Gamma_a\lambda^m\nonumber\\
  \delta\varphi^i&=&\bar\epsilon_m(\hat\Gamma^i)^m_{\ \ n}\lambda^n\nonumber\\
  \delta\lambda^m&=&\frac{1}{2}F_{ab}\Gamma^{ab}\epsilon^m
  +i\Gamma^a D_a\varphi^i(\hat\Gamma^i)^m_{\ \ n}\epsilon^n
  +\frac{i}{r}\varphi^i(\hat\Gamma^i\hat\Gamma^{34})^m_{\ \ n}\epsilon^n
  -i[\varphi^4,\varphi^5](\hat\Gamma^{45})^m_{\ n}\epsilon^n\nonumber\\
  &&+i\Gamma^5D_5(\varphi^3\hat\Gamma^3+\varphi^{i^\prime}\hat\Gamma^{i^\prime})^m_{\ n}\epsilon^n
  -i[\varphi^3,\varphi^{i^\prime}](\hat\Gamma^{3i^\prime})^m_{\ n}\epsilon^n
  -i[\varphi^1,\varphi^2](\hat\Gamma^{12})^m_{\ n}\epsilon^n
\end{eqnarray}
and
\begin{eqnarray}
  \delta A_5&=&-i\bar\epsilon_m\Gamma_5\psi^m\nonumber\\
  \delta\varphi^3&=&\bar\epsilon_m(\hat\Gamma^{3})^m_{\ \ n}\psi^n\ ,\ \
  \delta\varphi^{i^\prime}=\bar\epsilon_m(\hat\Gamma^{i^\prime})^m_{\ \ n}\psi^n\nonumber\\
  \delta\psi^m&=&F_{a5}\Gamma^{a5}\epsilon^m+i\Gamma^a D_a(\varphi^3\hat\Gamma^3
  +\varphi^{i^\prime}\hat\Gamma^{i^\prime})^m_{\ \ n}\epsilon^n
  +\frac{2i}{r}\varphi^{i^\prime}(\hat\Gamma^{i^\prime}\hat\Gamma^{34})^m_{\ \ n}\epsilon^n\nonumber\\
  &&+i\Gamma^5D_5\varphi^i(\hat\Gamma^i)^m_{\ \ n}\epsilon^n-i[\varphi^i,\varphi^3]
  (\hat\Gamma^{i3})^m_{\ \ n}\epsilon^n-i[\varphi^i,\varphi^{i^\prime}](\hat\Gamma^{ii^\prime}
  )^m_{\ \ n}\epsilon^n\ .
\end{eqnarray}
The on-shell supersymmetry algebra is given by
\begin{align}
	[\delta_{1},\delta_{2}]A_{\mu} &= 2i \xi^{\nu}\partial_{\nu}A_{\mu}+ 2i(\partial_{\mu}\xi^{\nu})A_{\nu}+\nabla_{\mu}\Lambda+i[\Lambda,A_{\mu}]=2i(\mathcal{L}_{\xi}A)_{\mu}+D_{\mu}\Lambda ,\nonumber \\
	[\delta_{1},\delta_{2}]\phi^{I} &= 2i \xi^{\mu}\partial_{\mu}\phi^{I}+i[\Lambda,\phi^{I}] - \frac{4}{r}(\bar\epsilon_{2}\hat{\Gamma}^{5}\epsilon_{1})\Big(-i(\delta^{I1}\delta^{J2}-\delta^{I2}\delta^{J1})\Big)\phi^{J}\;,
\end{align}
for the bosonic fields, where
\begin{eqnarray}
  \xi^{\mu} &=& \bar{\epsilon}_{2}\Gamma^{\mu}\epsilon_{1}\\
  \Lambda &=& -2i(\bar{\epsilon}_{2}\Gamma^{\mu}\epsilon_{1})A_{\mu} + 2(\bar{\epsilon}_{2}\hat{\Gamma}^{I}\epsilon_{1})\phi^{I} = -2i(\bar{\epsilon}_{2}\Gamma^{a}\epsilon_{1})A_{a} + 2(\bar{\epsilon}_{2}\hat{\Gamma}^{i}\epsilon_{1})\phi^{i}\nonumber
\end{eqnarray}
with $i=4,5$ and $a=1,2,3,4$. This shows that the algebra is $OSp(2|4)$, up to
a gauge transformation with parameter $\Lambda$. The algebra on fermionic fields should be
\begin{equation}
  [\delta_1,\delta_2]\Psi^m=2i\xi^\mu\partial_\mu\Psi^m
  +\frac{i}{2}\Theta^{ab}\Gamma_{ab}\Psi^m
  +i[\Lambda,\Psi^m]-\frac{4}{r}(\bar\epsilon_2\hat\Gamma^5\epsilon_1)
\left(-\frac{i}{2}(\hat\Gamma^{12})^m_{\ n}\right)\Psi^n\ ,
\end{equation}
where $\Theta^{ab}=\nabla^{[a}\xi^{b]}+\xi^{\lambda}\omega^{ab}_\lambda$ with the spin connection
$\omega^{ab}_\mu$ on $S^4$, which we have not checked. The $SO(2)_R$ R-symmetry rotates
$\varphi^1$ and $\varphi^2$ and leaves $\varphi^{3,4,5}$ invariant. Note that,
in generic 4d $\mathcal{N}=2^\ast$ on $S^4$ \cite{Pestun:2007rz}, $SO(2)_R$ rotates
$\varphi^{1,2}$ and also $A_5,\varphi^3$. However, at the special value $m=\pm\frac{1}{r}$
of hypermultiplet mass, it rotates $\varphi^{1,2}$ only, consistent with what we record
here (for $m=\frac{1}{r}$). Also, the Killing vector $\xi^\mu$ appearing
on the right hand side of the algebra only acts on $S^4$ part, i.e. $\xi^5=0$,
generating the $Sp(4)=SO(5)$ rotation on $S^4$.

The theory we found indeed has the correct reduction to the $\mathcal{N}=2^\ast$
theory on $S^4$ with special hypermultiplet mass $m=\frac{1}{r}$. See the appendix.
A simple but important aspect one can check from (\ref{action}) is
the scalar mass. The scalar $\varphi^3$ is massless, which combines with $A_5$ to
form two of the four hypermultiplet scalars. The remaining four scalars have net mass
$m_{\rm net}^2=\frac{2}{r^2}$: two of them are the other two scalars in the hypermultiplet,
while the remaining two are from the vector multiplet. This is exactly what we saw at the
beginning of this section.

The values for the 4d hypermultiplet mass parameter which allow the 5d uplifts
are $m=\pm\frac{1}{r}$, where the two signs are obtained by suitably changing the signs
of the backgrounds $S,V_5$. On round $S^4$, this corresponds to $m=\pm\epsilon_+$ at the north
and south poles of $S^4$ in the sense of \cite{Pestun:2007rz}, where
$\epsilon_+=\frac{\epsilon_1+\epsilon_2}{2}$ is the effective Omega deformation parameter in
the self-dual part near the poles. We shall see in the next subsection that the
$\mathcal{N}=2^\ast$ theory with general hypermultiplet mass uplifts to SYM on $S^4\times S^1$
with a defect on $S^1$.

The key requirement that the 4d theory should have massless scalars to admit an
uplift to the SYM on $S^4\times S^1$, and thus on $S^4\times\mathbb{R}$, is an essential
condition for the 6d background for the $(2,0)$ theory. To see the power of this
constraint, one can go to the squashed $S^4$ and apply the same logic. The study
of \cite{Hama:2012bg} on squashed $S^4$ is based on their metric and Killing spinor ansatz.
In particular, the metric is that on flat $\mathbb{R}^5$ induced on the following ellipsoid:
\begin{equation}
  \frac{x_0^2}{r^2}+\frac{x_1^2+x_2^2}{\ell^2}+\frac{x_3^2+x_4^2}{\tilde\ell^2}=1\ .
\end{equation}
Incidently, the analysis of \cite{Hama:2012bg} left three real functions $c_1,c_2,c_3$ of $S^4$
undetermined. Demanding that there exist two massless scalars in 4d, we find that $c_2,c_3$ are
algebraically determined, and $c_1$ is required to satisfy a complicated partial differential
equation. Thus, these functions are completely constrained, at least locally. Even with generic
4d hypermultiplet mass parameter, which is realized as the mass of 4d hypers on a defect, the possibility
of the 5d uplift would still constrain (and locally determine) the background.
We have not solved  these conditions in full generality. In a simple case with
$\ell=\tilde\ell$, the metric has $SO(4)$ isometry. In this case, we explicitly found
the globally regular solution
\begin{equation}
  c_1=-\frac{3}{4}\left(\frac{1}{\ell}-\frac{1}{\sqrt{r^2\sin^2\rho+\ell^2\cos^2\rho}}
  \right)\cot\rho\ ,\ \ c_2=0\ ,\ \ c_3=0\ ,
\end{equation}
admitting two massless scalars at $m=\pm\frac{1}{\ell}$.\footnote{This does not agree with the
exact $\Omega$-background of \cite{Hama:2012bg} around the north pole $\rho=0$,
presented in pp.14-15 there. Due to different $\rho$ scalings of the chiral and anti-chiral
Killing spinors of eqn.(3.40) there, we observe that the asymptotic form of some background
fields near north pole may have a finite deviation from the exact $\Omega$-background. The
finite deviations are suppressed by a factor of $\rho$ in the Killing spinor equation,
multiplied by the chiral Killing spinor $\xi_{\alpha A}\sim\mathcal{O}(\rho^1)$. It is
unclear to us whether such a deviation will affect the partition function calculus of
\cite{Hama:2012bg}. It deserves further studies.} $\rho$ is a coordinate of $S^4$
\cite{Hama:2012bg}, satisfying $0\leq\rho\leq\pi$. It will be interesting to generalize
this to the case with $\ell\neq\tilde\ell$.

\subsection{5d uplifts of more general 4d SYM}

We shall now discuss the 5d uplift of the $\mathcal{N}=2^\ast$ theory
with general 4d hypermultiplet mass. Since the existence of a massless hypermultiplet
scalar in 4d was crucial, we cannot uplift the
hypermultiplet with general mass into 5d fields. The 4d hypermultiplet with general mass
should come from degrees of freedom living on a 4d defect, transverse to the uplifted circle.
Note that \cite{Gaiotto:2009we} engineers the 4d $\mathcal{N}=2^\ast$
theory with general hypermultiplet mass by compactifying the 6d $(2,0)$ theory
on $T^2$ with a simple puncture. This comes from an intersecting M5-brane system,
whose type IIA reduction along a circle is the D4-NS5 system \cite{Witten:1997sc}.
The puncture of \cite{Gaiotto:2009we} reduces to the boundary of D4-branes ending on NS5,
on which a 4d hypermultiplet can live. The 5d theory on $S^4\times S^1$ with
a defect can be understood as living on $S^4\times I$, where $I$ is an interval of length
$2\pi r_1$, with suitable boundary conditions at the two ends. This theory has the flat space
limit $r\rightarrow\infty$ on $\mathbb{R}^4\times I$ with boundary degrees, which can be well
understood with the results of \cite{Gaiotto:2008sa}.
The SYM on $S^4\times I$ can in fact be understood as a building block of the `5d uplift'
of a larger class of gauge theories on $S^4$, obtained by wrapping M5-branes on Riemann surfaces, in the limit in which the Riemann surface degenerates \cite{Gaiotto:2009we}. These 5d SYM
coupled to boundaries may be a useful set-up to study the physics of M5-branes on
$S^4\times\Sigma_2$, possibly with instanton corrections.

We first explain the familiar boundary conditions on $\mathbb{R}^4\times I$, and then
elaborate on the case with $S^4\times I$.
We start by considering the brane realization of this SYM on flat space. This is given by
the NS5-D4 configuration of \cite{Witten:1997sc}, where NS5's are extended along $012345$,
and $N$ D4's are extended along $01236$ in $\mathbb{R}^{9,1}$. The $6$ direction is put on a
segment $I$, and a D4-brane ends on an NS5-brane at each end, with the boundary $\mathbb{R}^{3,1}$
along the $0123$ directions. Across a boundary of $I$, we put another set of $N$ D4-branes
starting from the same NS5-brane, also extended along $01236$. The relative displacement
of the two sets of $N$ D4-branes along the $45$ directions is labeled by a complex number
$\sim m$. The open strings ending on these two points provide a 4d bi-fundamental hypermultiplet
field with mass $m$. This field is supported on the `NS5-brane defect' localized in the $6$ direction. This way, we can form linear or circular quiver gauge theories in the 4d limit
\cite{Witten:1997sc}. The corresponding configurations of \cite{Gaiotto:2009we} are either
$N$ M5-branes wrapped on a sphere with $2$ full punctures and many simple punctures, or
$N$ M5-branes wrapped on a torus with many simple punctures.

Let us first summarize the boundary condition for D4-branes ending on
an NS5-brane, before taking the 4d boundary degrees into account. The 5d fields
should satisfy the following boundary conditions at an end of the interval.
For bosonic fields, they are
\begin{equation}
  \left.F_{a5}\frac{}{}\right|_{y=0}=0\ ,\ \
  \left.D_y\varphi^{4,5}\frac{}{}\right|_{y=0}=0\ ,\ \
  \left.\varphi^{1,2,3}\frac{}{}\right|_{y=0}=0\ .
\end{equation}
There are projection conditions for fermions as well. The $1,2,3$ directions
for the scalars denote the three directions transverse to the NS5-brane. $y\equiv x^5$ is
the coordinate for the interval, and $a=1,2,3,4$ is for the remaining $4$ directions. Such boundary conditions are imposed at the two ends of $I$,
say at $y=0,\beta$. Since $\varphi^{1,2,3}$ are constrained to be zero at the
two ends of the interval, the 4d masses for these 5d fields are all proportional to $\beta^{-1}$,
which become very heavy on a short interval and decouple. The $A_5$ field can also be set to
$0$ by using $y$ dependent local gauge transformation (where the gauge function is unconstrained
at the two boundaries).
Thus, all the four fields $\varphi^{1,2,3},A_5$ are set to zero in the 4d limit.
With 4d boundary degrees, the fields with Dirichlet boundary conditions will
satisfy modified Dirichlet boundary conditions \cite{Gaiotto:2008sa}.
However, the argument on the decoupling of the bulk fields on a short $I$ will remain
unchanged with the boundary degrees turned on (also with curvature corrections
on $S^4$). The boundary degrees will provide the hypermultiplet on $S^4$ in the 4d
limit with general mass.

The hypermultiplet that we introduce at the boundary of the interval
couples to the bulk 5d gauge fields in the following way.
Let us put the defect at $x^5=0$. There are two boundary values of the
fields $A_a,\lambda,\varphi^{4,5}$ which are subject to Neumann boundary conditions,
living on the interval on the right side $x^5>0$ and on the left $x^5<0$.
One of these two intervals may be semi-infinite. Let us call these two boundary values
as $A^\pm_a,\lambda^\pm,(\varphi^{4,5})^\pm$, respectively. Then the boundary
hypermultiplet would naively appear to be coupling to these the bulk fields in
the bi-fumdanental representation of $U(N)\times U(N)$. Of course we are
able to construct the 5d SYM coupling with the defect degrees in this way. However, there is
a subtle point on this gauge coupling \cite{Witten:1997sc}, if one wishes to realize the QFT
for the D4-NS5 system. Let us start by considering $(\varphi^{4,5})^{\pm}$, which represent
the end points of the D4-branes at the NS5-brane from the two sides. From the NS5-brane
dynamics, it was shown \cite{Witten:1997sc} that the modes with finite NS5-brane inertia
should satisfy
\begin{equation}
  \partial_a\left[{\rm tr}(\varphi^{4,5})^+-{\rm tr}(\varphi^{4,5})^-\right]=0
\end{equation}
at $x^5=0$. Extending this result to the full vector multiplet,
the dynamics of the relative $U(1)$ of $U(N)\times U(N)$ is frozen.
As other fields in the relative $U(1)$ is frozen to zero, only the constant (non-dynamical)
value of ${\rm tr}(\varphi^{4,5})^+-{\rm tr}(\varphi^{4,5})^-$ couples to the 4d degrees.
This is the mass $m$ of the hypermultiplet \cite{Witten:1997sc}. Thus,
only the $SU(N)\times SU(N)$ gauge fields dynamically couple to the 4d degrees,
since the overall $U(1)$ of $U(N)\times U(N)$ also decouples.

Now we explain the SYM on $S^4\times I$ with boundary degrees.
The boundary hypermultiplet action on $S^4$ with gauge coupling is completely dictated by
the analysis of \cite{Hama:2012bg}.\footnote{The formalism of \cite{Hama:2012bg} technically
requires the gauge group to be embedded in $Sp(r)$ with a suitable $r$. We simply wrote down
our SUSY action essentially using the results of \cite{Hama:2012bg}, but checked the
SUSY invariance of the coupled 5d-4d action independently.} The bulk action on $S^4\times S^1$
that we constructed in the previous subsection also has to be replaced by an action on
$S^4\times I$ with an interval $I$. The boundary terms for the bulk fields should also be
introduced. All such boundary
terms in the flat space limit can be taken from \cite{Gaiotto:2008sa},
using the formalism of 4d infinite dimensional gauge theory for the 5d SYM, and
the corresponding `4d D-term' fields. \cite{Gaiotto:2008sa}
in fact uses the 3d infinite dimensional gauge theory for the 4d bulk fields coupling to
the 3d boundary, but the same method can be applied to our 5d-4d system. There
are curvature corrections for the surface terms, which we justify by a brutal SUSY check.

To write down the coupled 5d-4d system, it is helpful to write all spinors
(matters, SUSY) in a way to make the $10=4+6$ dimensional decomposition clear.
These are summarized in the appendix. Firstly, the 4d action for the defect hypermultiplet
$q_A,\psi$ is given by \cite{Hama:2012bg}
\begin{eqnarray}\label{boundary-action}
  \hspace*{-1cm}S_{\rm 4d}\!&\!=\!&\!\int_{S^4}\!\!d^4x\sqrt{g}\
  {\rm tr}\left[\frac{}{}\!\!\right.D_a\bar{q}^AD^aq_A+\frac{2}{r^2}\bar{q}^Aq_A
  +\frac{m}{r}(\tau^3)^A_{\ \ B}\bar{q}^Bq_A-D^{I+}(\tau^I)^A_{\ \ B}q_A\bar{q}^B
  +D^{I-}q_A(\tau^I)^A_{\ \ B}\bar{q}^Bq_A\\
  \hspace*{-1cm}&&+\left(\bar{q}^A\varphi^{4}_+\!-\!\varphi^{4}_-\bar{q}^A\right)
  \left(\varphi^{4}_+q_A\!-\!q_A\varphi^{4}_-\right)
  +\left(\bar{q}^A\varphi^{5}_+\!-\!\varphi^{5}_-\bar{q}^A\!-\!im\bar{q}^A\right)
  \left(\varphi^{5}_+q_A\!-\!q_A\varphi^{5}_-\!-\!imq_A\right)+i\bar\psi\gamma^aD_a\psi
  \nonumber\\
  \hspace*{-1cm}&&
  +i\bar\psi\left(\varphi^5_+\psi-\psi\varphi^5_--im\psi\right)+\bar\psi\gamma^5
  \left(\varphi^4_+\psi-\psi\varphi^4_-\right)+\sqrt{2}\bar\psi(\lambda^A_+q_A-q_A\lambda^A_-)
  -\sqrt{2}\left(\bar{q}^A\bar\lambda_{A+}-\bar\lambda_{A-}\bar{q}^A\right)\psi
  \left.\!\!\frac{}{}\right]\nonumber
\end{eqnarray}
where $A,B=1,2$ are for $SU(2)_R$ (broken to $U(1)$ on $S^4$),
$\lambda_{A\pm}$ are boundary values of the 5d gaugino satisfying a symplectic-Majorana
condition as explained in the appendix. We took $\psi$ to be a Dirac fermion.
$D^I_\pm$ for $I=1,2,3$ are the boundary values of the bulk D-term auxiliary fields, which
we shall introduce shortly. $\tau^I$ are three Pauli matrices. In \cite{Hama:2012bg}, all terms
containing $m$ can be introduced by coupling the hypermultiplet to a background 4d vector
multiplet $\phi_m,\bar\phi_m$, $D_m^I$ (namely, eqn.(4.6) of \cite{Hama:2012bg})
for the $U(1)_F$ flavor symmetry on $(q_A,\psi)$. The full SUSY transformation for these fields will be explained below, after we explain the bulk action.
The above 4d action is the form in which the boundary fields couple to the $U(N)\times U(N)$
gauge fields. In case one restricts the 4d fields to couple only to the $SU(N)\times SU(N)$
part, one should replace all the 5d bulk fields by their traceless parts.
For instance, one should replace
\begin{equation}
  -D^{I+}(\tau^I)^A_{\ \ B}q_A\bar{q}^B
  +D^{I-}(\tau^I)^A_{\ \ B}\bar{q}^Bq_A\rightarrow
  -D^{I+}\left[(\tau^I)^A_{\ \ B}q_A\bar{q}^B-({\rm trace})\right]
  +D^{I-}\left[(\tau^I)^A_{\ \ B}\bar{q}^Bq_A-({\rm trace})\right]\ .\nonumber
\end{equation}
In case the 4d fields live at the intersection of a finite interval and a semi-infinite
region, one of the two bulk fields is taken to be nondynamical.
If one considers many 5d SYM on $S^4\times I$ connected to others in a quiver,
there should be many boundary actions of the form (\ref{boundary-action}).

Now we turn to the 5d action. We shall write the 5d bulk action plus extra boundary terms while
keeping the auxiliary $D^I$ fields. This makes up an off-shell vector multiplet in the 4d
sense, with $A_a$, $\lambda_A$, $\varphi^{4,5}$. The analysis below follows \cite{Gaiotto:2008sa} (SYM with boundaries on flat space), although we had to check SUSY ourselves to decide the
surface term at $\frac{1}{r}$ order. The 5d SYM action on $S^4\times I$ with two boundaries
at $y=y_1,y_2$ is given by
\begin{eqnarray}
  S_{\rm 5d}&=&\frac{1}{g_{YM}^2}\int_{S^4\times I}
  d^5x\sqrt{g}\ {\rm tr}\left[\frac{}{}\!\!\right.
  \frac{1}{4}F_{ab}^2+\frac{1}{2}F_{a5}^2+\frac{1}{2}(D_a\varphi^{i})^2
  +\frac{1}{2}(D_y\varphi^{i})^2-\frac{1}{2}[\varphi^4,\varphi^5]^2-\frac{1}{2}[\varphi^i,\varphi^I]^2
  \nonumber\\
  &&-\frac{1}{2}D^ID^I+D^I\left(D_y\varphi^I+\frac{i}{2}\epsilon^{IJK}[\varphi^J,\varphi^K]
  +\delta(y-y_1)\varphi^I(y_1)-\delta(y-y_2)\varphi^I(y_2)\right)\nonumber\\
  &&+\frac{1}{r^2}(\varphi^i)^2+\frac{1}{r^2}\left((\varphi^1)^2+(\varphi^2)^2\right)
  +\frac{2i}{r}\varphi^3D_y\varphi^5-\frac{2}{r}\varphi^5[\varphi^1,\varphi^2]
  +\frac{i}{2}\bar\lambda_A\gamma^\mu D_\mu\lambda^A
  +\frac{i}{2}\bar\chi_A\gamma^\mu D_\mu\chi^A\nonumber\\
  &&+\bar\chi_A D_y\lambda^A+\frac{i}{2}\bar\lambda_A[\varphi^5,\lambda^A]
  +\frac{1}{2}\chi_A\left(-i[\varphi^5,\chi^A]+\frac{1}{r}(\tau^3)^A_{\ B}\chi^B\right)
  +\frac{1}{2}\bar\lambda_A\gamma^5[\varphi^4,\lambda^A]
  +\frac{1}{2}\bar\chi_A\gamma^5[\varphi^4,\chi^A]\nonumber\\
  &&-\frac{1}{2}\bar\chi_A(\tau^I)^A_{\ B}[\varphi^I,\lambda^B]
  +\frac{1}{2}\bar\lambda_A(\tau^I)^A_{\ B}[\varphi^I,\chi^B]\left.\frac{}{}\!\!\right]
\end{eqnarray}
with $I=1,2,3$, $i=4,5$. After integrating out $D^I$, this is the SYM action on
$S^4\times\mathbb{R}$ we wrote down in section 2.1, up to surface terms.
Note that the term $-\frac{2i}{r}\varphi^5D_y\varphi^3$ we wrote in our SYM action
in the previous subsection is changed to $+\frac{2i}{r}\varphi^3D_y\varphi^5$ on the
third line: in other words, we have to add a surface term at $\frac{1}{r}$ order.

The actions $S_{\rm 5d}$ and $S_{\rm 4d}$ are separately invariant under the following
SUSY transformations:
\begin{eqnarray}
  \delta A_a&=&-i\bar\epsilon_A\gamma_a\lambda^A\\
  \delta\varphi^4&=&-i\bar\epsilon_A\gamma^5\lambda^A\ ,\ \ \delta\varphi^5=\bar\epsilon_A\lambda^A
  \nonumber\\
  \delta\lambda^A&=&\frac{1}{2}F_{ab}\gamma^{ab}\epsilon^A
  +(iD_a\varphi^5+D_a\varphi^4\gamma^5)\gamma^a\epsilon^A
  +[\varphi^4,\varphi^5]\gamma^5\epsilon^A\nonumber\\
  &&-iD^I(\tau^I)^A_{\ \ B}\epsilon^B+\frac{i}{r}(\varphi^4\gamma^5-i\varphi^5)
  (\tau^3)^A_{\ B}\epsilon^B\nonumber\\
  \delta D^I&=&-\bar\epsilon_A(\tau^I)^A_{\ B}\gamma^a D_a\lambda^B+i\bar\epsilon_A
  (\tau^I)^A_{\ B}\left(\gamma^5[\varphi^4,\lambda^B]+i[\varphi^5,\lambda^B]\right)
\end{eqnarray}
for the bulk `vector multiplet' fields (with $a,b=1,2,3,4$),
\begin{eqnarray}
  \delta A_5&=&-\bar\epsilon_A\chi^A\ \ ,\ \ \delta\varphi^I=-i\bar\epsilon_A(\tau^I)^A_{\ B}\chi^B\\
  \delta\chi^A&=&-iF_{ay}\gamma^a\epsilon^A-D_a\varphi^I(\tau^I)^A_{\ B}\gamma^a\epsilon^B
  +(iD_y\varphi^4\gamma^5-D_y\varphi^5)\epsilon^A\nonumber\\
  &&+\left(i[\varphi^4,\varphi^I]\gamma^5-[\varphi^5,\varphi^I]\right)(\tau^I)^A_{\ B}\epsilon^B
  +\frac{2}{r}\epsilon^{3IJ}\varphi^I(\tau^J)^A_{\ B}\epsilon^B\nonumber\\
  \delta\bar\chi_A&=&-i\bar\epsilon_A\gamma^a F_{ay}+\bar\epsilon_B\gamma^a(\tau^I)^B_{\ A}
  D_a\varphi^I+\bar\epsilon_A(i\gamma^5 D_y\varphi^4-D_y\varphi^5)\nonumber\\
  &&+\bar\epsilon_B(\tau^I)^B_{\ A}\left(-i\gamma^5[\varphi^4,\varphi^I]
  +[\varphi^5,\varphi^I]\right)+\frac{2}{r}\epsilon^{3IJ}\bar\epsilon_B(\tau^I)^B_{\ A}\varphi^J\nonumber
\end{eqnarray}
for the bulk `hypermultiplet' fields (with $I,J=1,2,3$), and
\begin{eqnarray}
  \hspace*{-.5cm}\delta q_A\!&\!=\!&\!\!-\sqrt{2}i\bar\epsilon_A\psi\ ,\ \ \delta\bar{q}^A=-\sqrt{2}i\bar\psi\epsilon^A\\
  \hspace*{-.5cm}\delta\psi\!&\!=\!&\!\!-\sqrt{2}D_a q_A\gamma^a\epsilon^A+
  \sqrt{2}i\left(\varphi^{4+}q_A\!-\!q_A\varphi^{4-}\right)\gamma^5\epsilon^A
  -\sqrt{2}\left(\varphi^{5+}q_A\!-\!q_A\varphi^{5-}\!-\!imq_A\right)\epsilon^A
  +\frac{\sqrt{2}i}{r}q_A(\tau^I)^A_{\ B}\epsilon^B\nonumber\\
  \hspace*{-.5cm}\delta\bar\psi\!&\!=\!&\!\!\sqrt{2}\bar\epsilon_A\gamma^a D_a \bar{q}^A
  -\sqrt{2}i\bar\epsilon_A\gamma^5\left(\varphi^{4-}\bar{q}^A\!-\!\bar{q}^A\varphi^{4+}\right)
  +\sqrt{2}\bar\epsilon_A\left(\varphi^{5-}\bar{q}^A\!-\!\bar{q}^A\varphi^{5+}\!+\!im\bar{q}^A\right)
  +\frac{\sqrt{2}i}{r}\bar\epsilon_B(\tau^I)^B_{\ A}\bar{q}^A\nonumber
\end{eqnarray}
for the boundary hypermultiplet fields.

The bulk action in the flat space limit $\frac{1}{r}\rightarrow 0$ can be
naturally understood by regarding the 5d gauge theory as a 4d gauge theory with `infinite
dimensional gauge group,' following \cite{Gaiotto:2008sa}. Namely, one regards the 5d
$y\equiv x^5$ dependent gauge transformation with finite gauge group
as a 4d gauge transformation with infinite dimensional gauge group.
\cite{Gaiotto:2008sa} applied this idea to the 4d maximal SYM theory with 3d boundary,
but it extends to our problem in one higher dimension.
As a warming up, following \cite{Gaiotto:2008sa}, let us rewrite the
bosonic part of the bulk hypermultiplet potential as
\begin{equation}\label{boundary-square}
  \frac{1}{2}(D_y\varphi^I)-\frac{1}{4}[\varphi^I,\varphi^I]^2=
  \frac{1}{2}\left(D_y\varphi^I+\frac{i}{2}\epsilon^{IJK}[\varphi^J,\varphi^K]\right)^2
  -\frac{i}{6}\partial_y\left(\epsilon^{IJK}\varphi^I[\varphi^J,\varphi^K]\right)\ .
\end{equation}
Note that $\frac{1}{2}(D_y\varphi^I)^2$, which is part of the 5d kinetic term,
is regarded in 4d viewpoint as part of the potential. The second term is
the boundary term which one can drop in the absence of boundaries. With a boundary,
only the first complete-square term should be kept in our action.
One can rewrite the first term as
\begin{equation}
  \frac{1}{2}\left(D_y\varphi^I+\frac{i}{2}\epsilon^{IJK}[\varphi^J,\varphi^K]\right)^2
  \ \rightarrow\ \ -\frac{1}{2}D^ID^I+D^I
  \left(D_y\varphi^I+\frac{i}{2}\epsilon^{IJK}[\varphi^J,\varphi^K]\right)
\end{equation}
by introducing three D-term fields, which can all be found in our action $S_{\rm 5d}$.
With boundaries, the on-shell value of $D^I$ from our action is given by
\begin{equation}
  D^I=D_y\varphi^I+\frac{i}{2}\epsilon^{IJK}[\varphi^J,\varphi^K]
  +\delta(y-y_1)\left(\varphi^I(y_1)-g_{YM}^2\mu^I_1\right)
  -\delta(y-y_2)\left(\varphi^I(y_2) -g_{YM}^2\mu^I_2\right)
\end{equation}
on the interval $y_1<y<y_2$, where
\begin{equation}
  \mu^I_1\equiv(\tau^I)^A_{\ \ B}\left[q_{(1)A}\bar{q}^B_{(1)}-\frac{1}{N}{\bf 1}_{N\times N}
  {\rm tr}(q_{(1)A}\bar{q}^B_{(1)})\right]\ ,\ \
  \mu^I_2\equiv(\tau^I)^A_{\ \ B}\left[\bar{q}^B_{(2)}q_{(2)A}-\frac{1}{N}{\bf 1}_{N\times N}
  {\rm tr}(\bar{q}^B_{(2)}q_{(2)A})\right]\ .
\end{equation}
$q_{(1)A}$, $q_{(2)A}$ are the boundary fields at $y=y_1,y_2$, respectively.
This is the hyper-Kahler moment map for the 4d infinite dimensional gauge group
in the presence of boundaries and boundary degrees \cite{Gaiotto:2008sa}. $\mu^I_{1,2}$
are the moment maps for the two $SU(N)$ gauge transformations acting on the boundary fields.

From the above action, we can understand the boundary conditions
for the bulk fields at $y=y_1,y_2$. The boundary values $D^I(y_1)\equiv D_1^I$,
$D^I(y_2)\equiv D^I_2$ appear linearly in the action from the surface terms
of $D^I$ contained in $S_{\rm 4d}$ and $S_{\rm 5d}$, since the bulk term $\int dy D^ID^I$
has extra infinitesimal factor $dy$ and can be ignored. So $D^I_{1,2}$ are Lagrange
multipliers, for the boundary conditions of the hypermultiplet scalars $\varphi^I$. They
are modifications of Dirichlet boundary conditions \cite{Gaiotto:2008sa},
\begin{equation}\label{modified-dirichlet}
  \varphi^I(y_1)=g_{YM}^2\mu^I_1\ ,\ \ \varphi^I(y_2)=g_{YM}^2\mu^I_2\ .
\end{equation}
The gauge field $A_y$ may be fixed to $0$ by using $y$ dependent gauge transformation
on the interval $I$, as explained before. Thus, the boundary values of bulk fields
$\varphi^I,A_y$ forming a hypermultiplet are all constrained in terms of the boundary degrees.
The boundary conditions for the bulk fields $A_a$, $\varphi^{4,5}$ forming 4d vector multiplet
can also be determined. In the flat space limit, they satisfy the
Neumann boundary conditions $F_{ay}=0$, $D_y\varphi^{4,5}=0$. Some of them are
modified in the presence of boundary degrees and curvature corrections.
Making a variation $\delta\varphi^{4,5}$ and demanding extremization of the action including
the surface terms, the modification for the $\varphi^{4,5}$ fields is given by
\begin{equation}\label{modified-neumann}
  \left[D_y\varphi^4\right]_{y=y_{1,2}}=\mp g_{YM}^2\frac{\delta S_{\rm 4d}}{\delta\varphi^4(y_{1,2})}
  \ ,\ \ \left[D_y\varphi^5\right]_{y=y_{1,2}}=\frac{2i}{r}\varphi^3(y_{1,2})
  \mp g_{YM}^2\frac{\delta S_{\rm 4d}}{\delta\varphi^5(y_{1,2})}\ ,
\end{equation}
where $\mp$ signs are for $y=y_1,y_2(>y_1)$, respectively.
The field $\varphi^3(y_{1,2})$ appearing on the right hand side
is $g_{YM}^2\mu^3_{1,2}$, from (\ref{modified-dirichlet}).

Let us focus on the 5d uplift of the $\mathcal{N}=2^\ast$ theory on $S^4$.
Here, the two ends of $I$ are coupled to the same boundary field, transforming
in the bi-fundamental representation of the bulk gauge field at $y=y_1=0$
and $y=y_2=2\pi r_1$. Here, $r_1$ is the circle radius if one views this system
as living on $S^4\times S^1$ with a defect at $y=0$. The mass $m$ for the
hypermultiplet in $S_{\rm  4d}$ is the twisted compactification parameter on $S^1$.
In the small circle limit, $r_1\rightarrow 0$, we have checked that the full
action reduces to the general $\mathcal{N}=2^\ast$ action with general mass $m$ on $S^4$.
Here we simply illustrate how this works with the bosonic action.
With given boundary fields $q_A$, the bulk fields $\varphi^I$ with $I=1,2,3$ satisfy
the modified Dirichlet boundary conditions. So the tower of higher Fourier modes
for these fields on $I$ become heavy with mass gap $\frac{1}{r_1}$ and decouple in the
small $r_1$ limit. More precisely, one can write
\begin{equation}
  \varphi^I=g_{YM}^2\mu^I_1(q)-g_{YM}^2(\mu^I_1-\mu^I_2)\frac{y}{2\pi r_1}+\cdots
  =g_{YM}^2\mu^I_1(q)-\frac{g_{YM}^2}{2\pi r_1}(\tau^I)^A_{\ \ B}[q_A,\bar{q}^B]y+\cdots\ ,
\end{equation}
with $0\leq y\leq 2\pi r_1$, where $\cdots$ denotes `higher modes' form a Fourier
expansion with nonzero wavenumbers on $I$. So at low energy, we ignore this
tower and the light mode of $\varphi^I$ is constrained by the 4d fields.
The coupling $-\frac{1}{2}D^ID^I+D^I\partial_y\varphi^I$ provides the required 4d D-term
potential for $q_A$ in the 4d limit:
\begin{equation}
  \frac{1}{g_{YM}^2}\int_0^{2\pi r_1}dy\left(-\frac{1}{2}D^ID^I+D^I\partial_y\varphi^I\right)
  \rightarrow-\frac{1}{2g_4^2}D^ID^I-D^I(\tau^I)^A_{\ B}[q_A,\bar{q}^B]\ .
\end{equation}
$g_4^2\equiv\frac{g_{YM}^2}{2\pi r_1}$ is the 4d gauge coupling.
One can also show that, with the above lowest mode, all
the other terms in $S_{\rm 5d}$ containing the bulk hyper fields $\varphi^I$,
$\chi$ do not contribute to the low energy action on $S^4$ with small $r_1$.
Moving on to the bulk
vector multiplet, the right hand sides of (\ref{modified-neumann}) all contain $r_1$
(with fixed 4d coupling $g_4^2=\frac{g_{YM}^2}{2\pi r_1}$) so that one recovers the
Neumann boundary conditions at both ends. So on a small $S^1$,
the lowest modes come from the zero modes of these fields on the interval. Thus the
5d bulk vector multiplet action reduces to the 4d vector multiplet action on $S^4$.
Combining this action with $S_{\rm 4d}$, we find that one obtains the
$\mathcal{N}=2^\ast$ theory on $S^4$ with general mass parameter $m$ \cite{Pestun:2007rz}.

So far, we discussed the gauge theories living on $S^4$ times many intervals,
$I_1,I_2,\cdots,I_n$, where the $n$ intervals either form a linear quiver or a
circular quiver. A 4d hypermultiplet in bi-fundamental representation connects
two different intervals, and fundamental 4d hypermultiplet couples to one end of
an interval. Another important ingredient of the 4d gauge theories of \cite{Gaiotto:2009we}
is the so-called $T_N$ theory, which has $SU(N)^3$ global symmetry. One may consider
coupling this $T_N$ theory to three 5d gauge theories at the end of the intervals.
Although we are quite ignorant on the microscopic description of this part for general $N$,
the case with $N=2$ would admit a Lagrangian description. Then the 6d $SU(2)$ theory
compactified on general Riemann surface would admit a `5d uplift' in the sense explained
in this subsection.

\section{Comments on the reduction on small $S^4$}

In this section, we briefly discuss the compactification of 5d SYM on a small $S^4$.
This setting could shed light on the AGT correspondence, maybe by exhibiting the effective
Liouville/Toda quantum mechanical description in this limit. In particular, a similar
problem of reducing the 6d $(2,0)$ theory on a small $S^3$ was shown to be very interesting
\cite{Cordova:2013cea}.

The energy scale of our interest is much smaller than $\frac{1}{r}$, where $r$ is the radius
of $S^4$. We would like to keep $\frac{1}{g_{YM}^2}\ll\frac{1}{r}$, so that the mass
of instanton particles is much lighter than the KK scale of $S^4$.
We are interested in the low energy effective quantum mechanics. There is an
obvious light degree, which is the s-wave of the massless scalar $\varphi^3(y)$ on $S^4$.
We find that other 5d fields do not provide any more light degrees, meaning that
all the modes carry nonzero frequencies proportional to $\frac{1}{r}$ on $\mathbb{R}$.
The effective quantum mechanical action for $\varphi^3(y)$ could receive perturbative
and non-perturbative corrections. We shall mostly speculate on what sort of ingredients
would be necessary to have the asserted Liouville/Toda physics.

Firstly, it is tempting to identify the light scalar $\varphi^3$ as the
variables of the Toda quantum mechanics. This is possible because the our quantum
mechanical system is gauged with $A_5(y)$. One can fix this gauge by diagonalizing
the real scalar $\varphi^3$. Among the $N$ eigenvalues, one of them corresponding to
the overall $U(1)$ decouples, yielding $N-1$ scalars which can possibly interact
with one another. The number of light degrees match with the number of variables in
the Toda mechanics. It is still unclear how the Toda potential could be
generated. However, accepting the above identification of the $N-1$ eigenvalues with
the Toda scalars, we consider how such a potential could possibly appear from
the 5d SYM viewpoint.

We first consider the 1d kinetic term obtained by classically reducing
the 5d SYM on a small $S^4$. The proper scaling limit is to keep the s-waves of $\varphi^3$
and $p\equiv\frac{\varphi^5}{r}$ finite in the small $S^4$ limit.
(Unlike $\varphi^5$, other massive modes simply decouple with $\varphi^3$ even after
similar scalings.) The mechanical action on Euclidean $\mathbb{R}$ is given by
\begin{equation}
  \frac{8\pi^2r^4/3}{g_{YM}^2}\int dy\
  {\rm tr}\left[\frac{1}{2}(D_y\varphi_3)^2+p^2+2i\varphi_3 D_y p\right]\ ,
\end{equation}
where the prefactor $\frac{8\pi^2r^4}{3}$ comes from the volume of $S^4$.
Without boundaries, we can integrate by part the last term and algebraically integrate out
the $p$ field to obtain
\begin{equation}
  \frac{4\pi^2r^4}{g_{YM}^2}\int dy\ {\rm tr}(D_y\varphi_3)^2\ .
\end{equation}
We fix the gauge symmetry by diagonalizing $\varphi^3$.
The 1-loop correction to the effective action from the heavy perturbative modes
on small $S^4$ should be computable in the background $\varphi^3(y)$. We have not
performed this computation, but this factor might cancel out or does not seriously
change (at least qualitatively) the nature of the above classical kinetic term.
We assume so in the considerations below, just to illustrate a possible (or hypothetical)
way of getting the Liouville-Toda potential from this approach.

Let us discuss the $U(2)$ theory for simplicity. Decomposing the overall
$U(1)$ and the rest by $\varphi_3=\varphi_0{\rm 1}_2+\frac{1}{2}\sigma^3\varphi$
the action for $\varphi$ is given by
\begin{equation}
  \frac{2\pi^2r^4}{g_{YM}^2}\int dy\ \dot\varphi^2\ .
\end{equation}
Thus, at this stage one obtains a free scalar action on $\mathbb{R}^+$,
after modding out by the Weyl gauge symmetry. Let us putatively interpret
this as the kinetic term of the Liouville action,
\begin{equation}\label{Liouville}
  \frac{1}{4\pi}\int d^2x\left(\partial^\mu\phi_L\partial_\mu\phi_L+
  4\pi\mu e^{2b\phi_L}\right)\ ,
\end{equation}
put on a cylinder and reduced on the small circle to mechanics.
$\phi_L$ denotes the scalar field in the above Liouville theory normalization.
Reducing the Liouville theory on a circle, one obtains the following quantum
mechanical action:
\begin{equation}
  \frac{g_{YM}^2}{8\pi^2}\int d\tau\left(\dot\phi_L^2+4\pi\mu e^{2b\phi_L}\right)
\end{equation}
where we write the circumference of the small circle as $2\pi r_2=\frac{g_{YM}^2}{2\pi}$,
interpreting this circle as the sixth circle which uplifts from the 5d SYM. (This relation
holds with our normalization for the Yang-Mills kinetic term
$\frac{1}{4g_{YM}^2}{\rm tr}(F_{\mu\nu}F^{\mu\nu})$.) From this, we make the following
identification of the Liouville scalar $\phi_L$ and the scalar $\varphi$ from the 5d SYM:
\begin{equation}
  \phi_L=\frac{4\pi^2r^2}{g_{YM}^2}\ \varphi\ .
\end{equation}
One can also rewrite the Liouville quantum mechanics action with our $\varphi$ variable.
Since we consider the round $S^4$, we insert $b=1$ in (\ref{Liouville})
and obtain
\begin{equation}
  \frac{2\pi^2r^4}{g_{YM}^2}\int dy\left(\dot\varphi^2+
  4\pi\tilde\mu e^{\frac{8\pi^2r^2}{g_{YM}^2}\varphi}\right)
\end{equation}
where $\mu\equiv\tilde\mu\frac{16\pi^4r^4}{g_{YM}^4}$. So the potential that is needed
for the Liouville quantum mechanics is
\begin{equation}\label{potential-SYM}
  \exp\left(4\pi r^2\cdot\frac{2\pi\varphi}{g_{YM}^2}\right)\ ,
\end{equation}
assuming our interpretation of $\varphi$ as $\phi_L$.

The potential takes the form of a non-perturbative correction
in the Weyl chamber $\varphi<0$.
So it would be interesting to think about what kind of non-perturbative effects could
account for (\ref{potential-SYM}) in the SYM on $S^4\times\mathbb{R}$. It is tempting
to make a somewhat wild speculation about (\ref{potential-SYM}). Namely, the prefactor
$4\pi r^2$ is the volume of a great 2-sphere cycle in $S^4$. So the above exponent could
be coming from a configuration wrapping this $S^2$, or a co-dimension $3$
finite action `instantons' on $S^4\times\mathbb{R}$ which is wrapping the $S^2$.
It is somewhat hard for us to imagine how such a finite action configuration could be
possible on $S^4\times\mathbb{R}$. Perhaps trying to reconsider an alternative
localization on $S^4$ might provide a hint, similar to \cite{Pestun:2009nn} which
manifestly keeps the $SO(3)$ isometry of the above $S^2$ factor.
Note also that, $\frac{2\pi|\varphi|}{g_3^2}$ is the action of a 't Hooft-Polyakov
monopole instanton in 3 dimensional gauge theory on $\mathbb{R}^3$ with gauge coupling
$g_3^2$ and scalar VEV $\varphi$ (again with our convention
$\mathcal{L}=\frac{1}{4g_{YM}^2}{\rm tr}\left(F_{\mu\nu}^2+\cdots\right)$ for
the Yang-Mills action).
So this makes us wonder whether a suitable stepwise compactification of the 5d SYM
to 3d and then to 1d would enable us to easily see the above non-perturbative effect.
For instance, considering the $S^4$ as a foliation of $S^2\times S^1$ over a segment
$0<\theta<\frac{\pi}{2}$ with metric
\begin{equation}
  ds_4^2=r^2\left(d\theta^2+\cos^2\theta ds^2(S^2)+\sin^2\theta d\psi^2\right)\ ,
\end{equation}
a formal reduction of the SYM on $S^2$ would yield
$\frac{1}{g_3^2}=\frac{4\pi r^2}{g_{YM}^2}$ near $\theta=0$. Presumably it should
be more appropriate to study the 5d SYM on highly squashed $S^4\times\mathbb{R}$,
by uplifting the gauge theory of \cite{Hama:2012bg} on squashed $S^4$ to 5d.
It would be interesting to see if these thoughts survive after
more rigorous investigations.

\section{SYM on $S^n\times\mathbb{R}$}

In this section, we discuss SYM theories on $S^n\times\mathbb{R}$.
Many such theories are known. For $n=2,3$, we shall simply summarize the theories
that are known or easily deducible from known results. For $n=4,5$, SYM on
$S^n$ provides a strong constraint and we only find SYM on $S^n\times\mathbb{R}$ with
the field content of maximal SYM. For $n\geq 6$, SYM is not allowed on
$S^n\times\mathbb{R}$ within our ideas. We start by summarizing known results.

On $S^3\times\mathbb{R}$, Yang-Mills action can be written down in the canonical way,
since it is classically conformal. Supersymmetric Yang-Mills theories can also
be written down easily. If the matter contents are suitably chosen, one can have an
$\mathcal{N}$-extended SCFT with $SU(2,2|\mathcal{N})$ symmetry at the quantum level.
We shall only discuss classical aspects of the superconformal action on $S^3\times \mathbb{R}$.
By suitably compactifying the theory on $S^1$, one can obtain SYM theories on $S^3$.
For simplicity, consider 4d $\mathcal{N}=1$ SCFT on $S^3\times\mathbb{R}$.
The 4d superconformal symmetry has $4$ Poincare SUSY $Q_\alpha$, $\bar{Q}_{\dot\alpha}$,
with $R=+1$ and $R=-1$ and $4$ conformal SUSY $S_\alpha$, $\bar{S}_{\dot\alpha}$ with
$R=-1$ and $R=+1$, respectively, where $R$ is the $U(1)$ R-charge. One can make a twisted
compactification on $S^1$ using $E-R/2$, where $E$ is the translation on $\mathbb{R}$
(dimension of operators). This compactification preserves half of the $8$ superconformal
symmetries which commute with $E-R/2$, namely $Q_{\alpha}$ and
$S_\alpha$. This should yield 3d $\mathcal{N}=2$ SYM with $OSp(2|2)$ symmetry, which were found in \cite{Jafferis:2010un,Hama:2010av}. The 3d theory has one real scalar $\sigma$ in the vector
multiplet, which comes from the holonomy of $A_4$ on $S^1$. From the 4d perspective,
$\sigma$ should be massless. This is in fact true, which one can check by integrating out
the D-term auxiliary field of \cite{Jafferis:2010un,Hama:2010av}.

Let us move on to $n=2$. For simplicity, we only consider the cases with $\mathcal{N}=(2,2)$
\cite{Benini:2012ui,Doroud:2012xw} or more SUSY. The 2d $(2,2)$ vector multiplet has two real
scalars, $\sigma_1$, $\sigma_2$. One scalar, say $\sigma_2$, is massless on $S^2$. Another scalar
$\sigma_1$ has the following coupling
\begin{equation}
  \left(F_{12}+\frac{1}{r}\sigma_1\right)^2\ ,
\end{equation}
where $F_{12}$ is the field strength in the frame basis. The presence of the massless scalar
$\sigma_2$ admits the possibility of an $S^2\times S^1$ uplift.
In fact, one can easily construct 3d $\mathcal{N}=2,4,8$ super-Yang-Mills theories consisting
of the vector multiplet. For $\mathcal{N}=8$, maximal SYM on $S^2\times\mathbb{R}$ is
known with $SU(2|4)$ symmetry \cite{Maldacena:2002rb,Lin:2005nh}. Starting from this,
one can obtain the
$\mathcal{N}=2,4$ truncations. Let us consider the case with $\mathcal{N}=2$ SUSY. The maximal
SYM has seven real scalars $X^a$, $\Phi$, four fermions $\Psi^{A}$, and complex Killing spinors
$\xi^A$, where $a=1,\cdots,6$ and $A=1,2,3,4$ for $SO(6)\sim SU(4)$ R-symmetry. We can consistently
turn off $X^a=0$ and $\Psi^{1,2,3}=0$, preserving $SU(2|1)$ symmetry. The fermionic symmetries
are parametrized by $\xi^4$. One can reduce this theory on $S^1$ preserving all $SU(2|1)$ SUSY,
by twisting $S^1$ translation $E$ by the $U(1)$ generator. As the complex SUSY $\xi^4$ has a
definite $U(1)$ charge, this twisting loses no SUSY and  yields the above
2d $\mathcal{N}=(2,2)$ theory, in which $\Phi=\sigma_1$, $A_3=\sigma_2$. We can also truncate
the maximal SYM to $\mathcal{N}=4$ SYM on $S^2\times\mathbb{R}$, by turning off $X^{1,2,3,4}=0$
and $\Psi^{1,2}=0$. One finds $SU(2|2)$ symmetry, whose fermionic generators are labeled by
$\xi^{3,4}$. The truncation can not be extended beyond $\mathcal{N}=4$, which should be the
case since there are no such theories even in the flat space limit. Coupling matters to these
$\mathcal{N}=2,4$ theories presumably should be possible, which we do not discuss.

Now let us move on to higher dimensions, $S^n\times\mathbb{R}$ or $S^n\times S^1$ with
$n\geq 5$. We first consider the case with $n=5$. On $S^5$ with radius $r$, the real
scalar in the $\mathcal{N}=1$ vector multiplet has mass $\frac{2}{r}$. So
one cannot uplift $\mathcal{N}=1$ SYM with vector multiplet only to $S^5\times S^1$.
However, like the SYM on $S^4$, uplift to $S^5\times S^1$ is possible with an
adjoint hypermultiplet. The bosonic action for the vector multiplet and
an adjoint hypermultiplet with mass $m$ is given by \cite{Hosomichi:2012ek,Kim:2012ava}
\begin{eqnarray}
  g_{YM}^2\mathcal{L}_{\rm bos}&=&{\rm tr}\left[\frac{}{}\!\!\right.
  \frac{1}{4}F_{\mu\nu}F^{\mu\nu}+\frac{1}{2}(D_\mu\phi)^2+|D_\mu q^A|^2+\frac{5}{2r^2}\phi^2
  +\frac{15}{4r^2}|q^A|^2-\frac{1}{2}D^ID^I-\frac{i}{r}\phi D^3\nonumber\\
  &&\hspace{.5cm}\left.+\left([\bar{q}_A,\phi]-im\bar{q}_A\right)\left([\phi,q^A]-imq^A\right)
  -\bar{q}_A(\tau^I)^A_{\ B}\left([D^I,q^B]+\delta^I_3\frac{m}{r}q^B\right)\right]\ ,
\end{eqnarray}
where $m=\frac{1}{r}\left(\Delta-\frac{1}{2}\right)$ in the notation of \cite{Kim:2012ava},
and $I=1,2,3$, $A,B=1,2$ for the $SU(2)_R$ symmetry broken to $U(1)_R$.
$\phi$ is the real scalar in the vector multiplet, and $q^1,q^2$ are the two complex scalars
in the hypermultiplet. With general $m$, this SYM preserves $SU(4|1)$ symmetry with $8$ SUSY.
After integrating out the auxiliary $D^I$ fields, the mass terms are given by
\begin{equation}
  \frac{2}{r^2}\phi^2+\bar{q}_A\left[\left(\frac{15}{4r^2}-m^2\right)\delta^A_{\ B}
  -\frac{m}{r}(\tau^3)^A_{\ B}\right]q^B\ .
\end{equation}
From this, one finds that one of $q^1,q^2$ becomes massless at $m=\pm\frac{3}{2r}$.
At these values, another complex scalar has net mass-square $\frac{3}{r^2}$, and the real
scalar $\phi$ has mass-square $\frac{4}{r^2}$. The 5d theory at these values of mass can be
uplifted to $S^5\times S^1$, with one of the two massless scalars uplifting to $A_6$. This
can be easily convinced by again relying on a deconstruction-like argument. The above SYM on
$S^5$ can be written down with arbitrary gauge group and matter content, so we consider the
$U(N)^K$ theory with $K$ bifundamental hypermultiplets forming a circular quiver. Although the
full quantum deconstruction like \cite{ArkaniHamed:2001ie} is not expected to exist, as both
6d and 5d theories are nonrenormalizable, one can still discuss it at the level of discretizing
higher dimensional classical field theory, in the spirit of \cite{Lambert:2012qy}.\footnote{From
brane perspective, $N$ Dp-branes probing $\mathbb{C}^2/\mathbb{Z}_K\times\mathbb{R}^{5-p}$ engineer $p+1$ dimensional circular quiver theory. Higgsing with large $K$, the
$\mathbb{C}/\mathbb{Z}_K$ part of the geometry probed by the Higgs branch would approximate
to $\mathbb{R}\times S^1$. T-duality along $S^1$ (equivalent to the Fourier transformation in
the deconstruction of \cite{Lambert:2012qy}) yields $p+2$ dimensional maximal SYM.} Taking all
$K$ hypermultiplet mass parameters to be, say $m=\frac{3}{2r}$, one can give Higgs VEV and
take large $K$ scaling limit like \cite{ArkaniHamed:2001ie,Lambert:2012qy} to provide massive
Kaluza-Klein modes on $S^1$. The full action on $S^5\times S^1$ or $S^5\times\mathbb{R}$ can
be obtained, although one has to pay some effort to convert the spinor convention to what is
more natural in 6d. We simply write the bosonic action here. Let us take $m=\frac{3}{2r}$, and
call $q^1=\frac{A_6-i\phi^3}{\sqrt{2}}$, $q^2=-\frac{\phi^2+i\phi^1}{\sqrt{2}}$.
Then the bosonic part of the 6d SYM action on $S^5\times\mathbb{R}$ is given by
\begin{eqnarray}
  g_{6}^2\mathcal{L}_{\rm bos}&=&{\rm tr}
  \left[\frac{}{}\!\!\right.\frac{1}{4}F_{\mu\nu}F^{\mu\nu}
  +\frac{1}{2}(F_{\mu y})^2+\frac{1}{2}(D_\mu\phi)^2+\frac{1}{2}(D_y\phi)^2
  +\frac{1}{2}(D_\mu\phi^I)^2-\frac{1}{2}[\phi,\phi^I]^2\nonumber\\
  &&-\frac{1}{2}D^ID^I+D^I\left(D_y\phi^I+\frac{i}{2}\epsilon^{IJK}[\phi^J,\phi^K]
  -\frac{i}{r}\phi\delta^I_3\right)+\frac{3i}{r}\phi\left(D_y\phi^3-i[\phi^1,\phi^2]\right)\nonumber\\
  &&+\frac{5}{2r^2}\phi^2+\frac{3}{2r^2}((\phi^1)^2+(\phi^2)^2)\left.\frac{}{}\!\!\right]\ ,
\end{eqnarray}
where $\mu,\nu=1,2,3,4,5$, $y\equiv x^6$ and $I=1,2,3$, and
$g_{6}$ is the 6d Yang-Mills coupling.

The studies of the 6d maximal SYM on $S^5\times\mathbb{R}$ or $S^5\times S^1$ may
be interesting in the context of type IIB  little string theory with $(1,1)$ SUSY.
The $S^5$ partition function
acquires contributions from three instanton partition functions on $\mathbb{R}^4\times S^1$
\cite{Lockhart:2012vp,Kim:2012qf,Qiu:2013aga}. Thus one could think that the $S^5\times S^1$
partition function would be obtained by combining three instanton partition functions
of 6d SYM on $\mathbb{R}^4\times T^2$ \cite{Hollowood:2003cv}, where the extra circle direction
comes from the $S^1$ uplift.
There appears one subtlety in this uplift, from the fact that one real scalar is massless
in 6d. The massless scalar will cause a divergence of the perturbative partition function
on $S^5$ as we take $m\rightarrow\frac{3}{2r}$. This divergence happens in the diagonal
$U(1)^N$ part of the perturbative partition function \cite{Kim:2012ava}. There will thus
appear a net $(mr-\frac{3}{2})^{-N}$ divergence. This is precisely the divergence coming from
the half-BPS partition function of the 6d $(2,0)$ theory, if one views the $S^5$ partition
function as the $(2,0)$ index. However, the residue of the partition function at $m=\frac{3}{2r}$
is finite. A simple calculation using the results of \cite{Kim:2012qf,Lockhart:2012vp} yields
\begin{equation}\label{S5-limit}
  Z_{S^5}\rightarrow\frac{1}{N!\beta^N(\frac{3}{2}-mr)^N}\cdot
  \frac{1}{\eta(e^{-\beta(1+a)})^N}\cdot\frac{1}{\eta(e^{-\beta(1+b)})^N}\cdot
  \frac{1}{\eta(e^{-\beta(1+c)})^N}
\end{equation}
in the $m\rightarrow\frac{3}{2r}$ limit, apart from the zero point energy factor.
Here $\beta=\frac{g_{YM}^2}{2\pi r}$ is the chemical potential for the `energy,'
and $\beta a$, $\beta b$, $\beta c$ are the chemical potential for the $SU(3)\subset SO(6)$
angular momentum on $S^5$: see \cite{Kim:2012qf} for the details. $\eta(q)$ is given by
$\eta(q)=q^{\frac{1}{24}}\prod_{n=1}^\infty(1-q^n)$.
The first factor is the $m\rightarrow\frac{3}{2r}$ limit of the $U(N)$
half-BPS partition function:
\begin{equation}
  \prod_{n=1}^N\frac{1}{1-e^{-n(\frac{3}{2}-mr)\beta}}\rightarrow\frac{1}
  {N!\beta^N(\frac{3}{2}-mr)^N}\ .
\end{equation}
The result (\ref{S5-limit}) is somewhat boring, as the residue at $mr=\frac{3}{2}$ just
takes the form of the $U(1)^N$ index. This is natural as this can be interpreted as
the IR index after Higgsing the theory with a complex scalar. It would be
more interesting to study the defects on $S^1$. For instance, the 5d version of the AGT
proposals and q-deformed CFT's studied in \cite{Nieri:2013yra} may be explored, if it
has a higher dimensional origin like \cite{Gaiotto:2009we,Alday:2009aq}.

Finally, at $n=6,7$, maximal SYM on $S^n$ is known in the literature
\cite{Blau:2000xg,Fujitsuka:2012wg}.\footnote{For $n=6$, the superalgebra should be $F(4)$
since it has $SO(7)\times SU(2)_R$ symmetry. For $n=7$, the superalgebra should be $OSp(8|2)$
since it has $SO(8)\times SU(2)_R$.} On $S^6$, one scalar have mass-square $\frac{4}{r^2}$,
and three have $\frac{6}{r^2}$. On $S^7$, the three scalars have mass-square $\frac{3}{r^2}$.
So there are no massless scalars in either case. It is also impossible to provide deformations
like extra hypermultiplet mass to have massless scalars. In 6d, hypermultiplet cannot be given
a mass parameter already in flat space limit, as the fermion of 6d hypermultiplet is chiral.
Also, there is no notion of hypermultiplet in 7d, and thus no way to tune the mass matrix.
So we cannot use our argument to have a SYM on $S^n\times\mathbb{R}$ at $n=6,7$. This seems to
lead to the conclusion that $n+1=6$ is the maximal dimension in which one can write down SYM
on $S^n\times\mathbb{R}$.

\vskip 0.5cm

\hspace*{-0.8cm} {\bf\large Acknowledgements}
\vskip 0.2cm

\hspace*{-0.75cm} We thank Kazuo Hosomichi, Hee-Cheol Kim, Costis Papageorgakis and
Shuichi Yokoyama for helpful discussions. This work is supported by the National Research Foundation of Korea (NRF) Grants No. 2012R1A1A2042474 (JK,SK), 2012R1A2A2A02046739 (SK),
2006-0093850 (KL), 2009-0084601 (KL), 2012-009117 (JP), 2012-046278 (JP). J.P. also
appreciates APCTP for its stimulating environment for research.

\appendix

\section{Spinor conventions}

As explained in (\ref{5d-10d-gamma}), we can conveniently uplift our $SO(5)\times SO(5)_R$
spinors into a $10$ dimensional spinor by using the following $32\times 32$ gamma matrices:
\begin{eqnarray}
  {\bf \Gamma}^\mu&=&\Gamma^\mu\otimes\hat\Gamma^5\otimes\sigma_1\\
  {\bf \Gamma}^{i+5}&=&{\bf 1}_4\otimes i\hat\Gamma^{5i}\otimes\sigma_1\ \ \ (i=1,2,3)\nonumber\\
  {\bf \Gamma}^9&=&{\bf 1}_4\otimes i\hat\Gamma^{54}\otimes\sigma_1\ ,\ \
  {\bf \Gamma}^0={\bf 1}_4\otimes{\bf 1}_4\otimes\sigma_2\ .\nonumber
\end{eqnarray}
Our convention for $\Gamma^\mu$ and $\hat\Gamma^I$ are explained in (\ref{SO(5)-gamma}) and
(\ref{SO(5)R-gamma}). The 5d spinor bilinears transform to
the following 10d bilinears,
\begin{equation}
  \bar\Psi_1\Gamma^\mu\Psi_2\rightarrow\bar{\bf \Psi}_1{\bf \Gamma}^\mu{\bf \Psi}_2\ ,\ \
  \bar\Psi_1\Gamma^I\Psi_2\rightarrow i\bar{\bf \Psi}_1{\bf \Gamma}^I{\bf \Psi}_2\ ,
\end{equation}
where ${\bf \Psi}_{1,2}$ are the corresponding 10d spinors satisfying the Weyl condition
$\sigma^3{\bf \Psi}_{1,2}={\bf \Psi}_{1,2}$. 10d barred spinors are defined by
$\bar{\bf \Psi}\equiv\Psi^\dag(-i{\bf \Gamma}^0)$, where $-i{\bf \Gamma}^0$ is the
`time component' of the Gamma matrix for the 10d theory in the $(9,1)$ signature theory
obtained by Wick rotation \cite{Pestun:2007rz}. The symplectic-Majorana condition for
the 5d spinors uplifts to
\begin{equation}\label{majorana-10d}
  \bar{\bf \Psi}={\bf \Psi}^TC_{10}\ ,\ \
  C_{10}\equiv C\otimes\Omega\hat\Gamma^5\otimes\sigma_1
  \ ,\ \ C_{10}({\bf \Gamma}^M)^TC_{10}^{-1}={\bf \Gamma}^M\ \ \ (M=1,2,\cdots,0)\ .
\end{equation}
Using the above bilinear relations and also using
\begin{equation}
  -\frac{i}{4r}\bar\Psi(\hat\Gamma^{34}+i\hat\Gamma^{35}\Gamma^5)\Psi\rightarrow\frac{1}{4r}
  \bar{\bf \Psi}{\bf \Gamma}^0\left({\bf \Gamma}^{67}+{\bf \Gamma}^{58}\right){\bf \Psi}\ ,
\end{equation}
one can show that our 5d action (\ref{action}) uplifts in the 10d notation to
\begin{eqnarray}
  S&=&\frac{1}{g_{YM}^2}
  \int d^5x\sqrt{g}\ {\rm tr}\left[\!\!\frac{}{}\right.\frac{1}{4}F_{MN}F^{MN}
  +\frac{i}{2}\bar{\bf \Psi}{\bf \Gamma}^MD_M{\bf \Psi}
  -\frac{2i}{r}\varphi^5\left(D_5\varphi^3\!-\!i[\varphi^1,\varphi^2]\right)\nonumber\\
  &&\hspace{1.5cm}
  +\frac{1}{r^2}\left((\varphi^1)^2\!+\!(\varphi^2)^2\!+\!(\varphi^4)^2\!+\!(\varphi^5)^2\right)
  +\frac{1}{4r}\bar{\bf \Psi}{\bf \Gamma}^0\left({\bf \Gamma}^{67}+{\bf \Gamma}^{58}\right){\bf \Psi}\left.\!\!\frac{}{}\right]
\end{eqnarray}
where $A_M\equiv(A_\mu,\varphi^I)$, $F_{IJ}\equiv-i[\varphi^I,\varphi^I]$,
$D_I{\bf \Psi}\equiv-i[\varphi^I,{\bf \Psi}]$, $F_{\mu I}\equiv D_\mu\varphi^I$.
Compactifying $x^5$ direction to a small circle and reducing to 4d,
one defines $\Phi^M=(A_5,\varphi^{I+5})$ with $P=5,\cdots,0$. Then the 4d action is
given by
\begin{eqnarray}
  &&\frac{1}{g_4^2}\int d^4x\sqrt{g}\ {\rm tr}
  \left[\frac{1}{4}F_{MN}F^{MN}-iD_0\Phi^i(M_{ij}\Phi^j)-\frac{1}{2}(M_{ij}\Phi^j)(M_{ik}\Phi^k)
  +\frac{1}{r^2}\Phi^P\Phi^P-\frac{1}{2r}R_{ki}M_{kj}\Phi^i\Phi^j\right.\nonumber\\
  &&\left.\hspace{3cm}
  +\frac{i}{2}\bar{\bf \Psi}{\bf \Gamma}^MD_M{\bf \Psi}+\frac{1}{8r}\bar{\bf\Psi}
  {\bf \Gamma}^0M_{ij}{\bf\Gamma}^{ij}{\bf \Psi}\ ,
  \right]
\end{eqnarray}
where $i,j=5,6,7,8$, and $R,M$ are given by
(\ref{pestun-matrices}) with $m=\frac{1}{r}$. This reproduces the special $\mathcal{N}=2^\ast$
action on $S^4$ with mass parameter $\frac{1}{r}$, where $D_0\Phi^i-iM_{ij}\Phi^j$ and
$\left(D_0-\frac{i}{4}M_{ij}{\bf\Gamma}^{ij}\right){\bf \Psi}$ combinations come from
the Scherk-Schwarz mass assignment.

To couple the 5d system to 4d boundary degrees in section 2.2, it is more
useful to assume the following $4+4+2$ decomposition of the 10d gamma matrices:
\begin{equation}
  {\bf \Gamma}^a=\gamma^a\otimes{\bf 1}_4\otimes{\bf 1}_2\ ,\ \
  {\bf \Gamma}^{4+i}=\gamma^5\otimes\hat\gamma^i\otimes\sigma_2\ ,\ \
  {\bf \Gamma}^9=\gamma^5\otimes{\bf 1}_4\otimes\sigma_1\ ,\ \
  {\bf \Gamma}^0=\gamma^5\otimes\hat\gamma^5\otimes\sigma^2\ ,
\end{equation}
with $a=1,2,3,4$, and we take
\begin{equation}
  \hat\gamma^i=\left(\begin{array}{cc}&(\sigma^i)_{A\dot{B}}\\
  (\bar\sigma^i)^{\dot{A}B}&\end{array}\right)\ ,\ \
  \hat\gamma^5=\left(\begin{array}{cc}-\delta_A^{\ B}&\\
  &\delta^{\dot{A}}_{\ \dot{B}}\end{array}\right)
\end{equation}
with $\sigma^i=(1,-i\vec\tau)$, $\bar\sigma^i=(1,i\vec\tau)$.
The projection ${\bf\Gamma}^{5678}\epsilon=\epsilon$ becomes $\hat\gamma^5\epsilon=\epsilon$,
meaning that $\epsilon^{\dot{A}}$ part generates 8 SUSY while $\epsilon_A$ is broken.
If we write the 5d action in this convention, such as the 5d SYM on $S^4\times I$ in
section 2.2, the fifth direction corresponding to
${\bf\Gamma}^5=\gamma^5\otimes\hat\gamma^1\otimes\sigma^2$ is picked. So the
$SO(4)$ rotation acting on the $i$ type indices breaks to $SO(3)$, even in the
flat space limit. Since this $SO(3)$ is the diagonal of the two $SU(2)$ rotations acting on
the $A,\dot{A}$ indices, the $A$ and $\dot{A}$ indices are identified. This is the $A$
doublet indices for $SU(2)_R$ that we use in section 2.2. Reduction of the 10d Majorana
condition yields the symplectic-Majorana condition in 4d, which is the one
used in \cite{Hama:2012bg}. This reality condition applies to our 5d spinors
$\lambda^A$, $\chi^A$ in section 2.2. In this spinor basis, the Killing spinor
equation for $\epsilon^A$ on $S^4$ is given by
\begin{equation}
  \nabla_a\epsilon^A=-\frac{i}{2r}\gamma_a(\tau^3)^A_{\ B}\epsilon^B\ .
\end{equation}


\begin{thebibliography}{12345}


\bibitem{Blau:2000xg}
  M.~Blau,
  ``Killing spinors and SYM on curved spaces,''
  JHEP {\bf 0011}, 023 (2000)
  [hep-th/0005098].

\bibitem{Benini:2012ui}
  F.~Benini and S.~Cremonesi,
  arXiv:1206.2356 [hep-th].

\bibitem{Doroud:2012xw}
  N.~Doroud, J.~Gomis, B.~Le Floch and S.~Lee,
  JHEP {\bf 1305}, 093 (2013)
  [arXiv:1206.2606 [hep-th]].


\bibitem{Jafferis:2010un}
  D.~L.~Jafferis,
  JHEP {\bf 1205}, 159 (2012)
  [arXiv:1012.3210 [hep-th]].

\bibitem{Hama:2010av}
  N.~Hama, K.~Hosomichi and S.~Lee,
  JHEP {\bf 1103}, 127 (2011)
  [arXiv:1012.3512 [hep-th]].

\bibitem{Pestun:2007rz}
  V.~Pestun,
  ``Localization of gauge theory on a four-sphere and supersymmetric Wilson loops,''
  Commun.\ Math.\ Phys.\  {\bf 313}, 71 (2012)
  [arXiv:0712.2824 [hep-th]].

\bibitem{Hosomichi:2012ek}
  K.~Hosomichi, R.~-K.~Seong and S.~Terashima,
  Nucl.\ Phys.\ B {\bf 865}, 376 (2012)
  [arXiv:1203.0371 [hep-th]].


\bibitem{Aharony:2003sx}
  O.~Aharony, J.~Marsano, S.~Minwalla, K.~Papadodimas and M.~Van Raamsdonk,
  Adv.\ Theor.\ Math.\ Phys.\  {\bf 8}, 603 (2004)
  [hep-th/0310285].


\bibitem{Romelsberger:2005eg}
  C.~Romelsberger,
  Nucl.\ Phys.\ B {\bf 747}, 329 (2006)
  [hep-th/0510060].

\bibitem{Kinney:2005ej}
  J.~Kinney, J.~M.~Maldacena, S.~Minwalla and S.~Raju,
  Commun.\ Math.\ Phys.\  {\bf 275}, 209 (2007)
  [hep-th/0510251].

\bibitem{Bhattacharya:2008zy}
  J.~Bhattacharya, S.~Bhattacharyya, S.~Minwalla and S.~Raju,
  JHEP {\bf 0802}, 064 (2008)
  [arXiv:0801.1435 [hep-th]].

\bibitem{Kim:2009wb}
  S.~Kim,
  Nucl.\ Phys.\ B {\bf 821}, 241 (2009)
  [Erratum-ibid.\ B {\bf 864}, 884 (2012)]
  [arXiv:0903.4172 [hep-th]].

\bibitem{Imamura:2011su}
  Y.~Imamura and S.~Yokoyama,
  JHEP {\bf 1104}, 007 (2011)
  [arXiv:1101.0557 [hep-th]].

\bibitem{Kim:2012gu}
  H.~-C.~Kim, S.~-S.~Kim and K.~Lee,
  JHEP {\bf 1210}, 142 (2012)
  [arXiv:1206.6781 [hep-th]].

\bibitem{Iqbal:2012xm}
  A.~Iqbal and C.~Vafa,
  arXiv:1210.3605 [hep-th].


\bibitem{Kim:2012ava}
  H.~-C.~Kim and S.~Kim,
  JHEP {\bf 1305}, 144 (2013)
  [arXiv:1206.6339 [hep-th]].

\bibitem{Kallen:2012zn}
  J.~Kallen, J.~A.~Minahan, A.~Nedelin and M.~Zabzine,
  JHEP {\bf 1210}, 184 (2012)
  [arXiv:1207.3763 [hep-th]];
  J.~A.~Minahan, A.~Nedelin and M.~Zabzine,
  J.\ Phys.\ A {\bf 46}, 355401 (2013)
  [arXiv:1304.1016 [hep-th]].

\bibitem{Kim:2012tr}
  H.~-C.~Kim and K.~Lee,
  JHEP {\bf 1307}, 072 (2013)
  [arXiv:1210.0853 [hep-th]].

\bibitem{Lockhart:2012vp}
  G.~Lockhart and C.~Vafa,
  arXiv:1210.5909 [hep-th].

\bibitem{Kim:2012qf}
  H.~-C.~Kim, J.~Kim and S.~Kim,
  arXiv:1211.0144 [hep-th].

\bibitem{Kim:2013nva}
  H.~-C.~Kim, S.~Kim, S.~-S.~Kim and K.~Lee,
  arXiv:1307.7660 [hep-th].

\bibitem{Gadde:2013dda}
  A.~Gadde and S.~Gukov,
  JHEP {\bf 1403}, 080 (2014)
  [arXiv:1305.0266 [hep-th]].

\bibitem{Benini:2013nda}
  F.~Benini, R.~Eager, K.~Hori and Y.~Tachikawa,
  Lett.\ Math.\ Phys.\  {\bf 104}, 465 (2014)
  [arXiv:1305.0533 [hep-th]];
  F.~Benini, R.~Eager, K.~Hori and Y.~Tachikawa,
  arXiv:1308.4896 [hep-th].


\bibitem{Witten:1995zh}
  E.~Witten,
  ``Some comments on string dynamics,''
  In *Los Angeles 1995, Future perspectives in string theory* 501-523
  [hep-th/9507121].

\bibitem{Strominger:1995ac}
  A.~Strominger,
  ``Open p-branes,''
  Phys.\ Lett.\ B {\bf 383}, 44 (1996)
  [hep-th/9512059].

\bibitem{Witten:1995em}
  E.~Witten,
  Nucl.\ Phys.\ B {\bf 463}, 383 (1996)
  [hep-th/9512219].



\bibitem{Douglas:2010iu}
  M.~R.~Douglas,
  JHEP {\bf 1102} (2011) 011;

\bibitem{Lambert:2010iw}
  N.~Lambert, C.~Papageorgakis and M.~Schmidt-Sommerfeld,
  JHEP {\bf 1101}, 083 (2011);

\bibitem{Bern:2012di}
  Z.~Bern, J.~J.~Carrasco, L.~J.~Dixon, M.~R.~Douglas, M.~von Hippel and H.~Johansson,
  Phys.\ Rev.\ D {\bf 87}, 025018 (2013)
  [arXiv:1210.7709 [hep-th]].

\bibitem{Papageorgakis:2014dma}
  C.~Papageorgakis and A.~B.~Royston,
  arXiv:1404.0016 [hep-th].


\bibitem{Kim:2011mv}
  H.~-C.~Kim, S.~Kim, E.~Koh, K.~Lee and S.~Lee,
  JHEP {\bf 1112}, 031 (2011)
  [arXiv:1110.2175 [hep-th]].

\bibitem{Haghighat:2013gba}
  B.~Haghighat, A.~Iqbal, C.~Kozcaz, G.~Lockhart and C.~Vafa,
  arXiv:1305.6322 [hep-th].


\bibitem{Alday:2009aq}
  L.~F.~Alday, D.~Gaiotto and Y.~Tachikawa,
  Lett.\ Math.\ Phys.\  {\bf 91}, 167 (2010)
  [arXiv:0906.3219 [hep-th]].

\bibitem{Wyllard:2009hg}
  N.~Wyllard,
  JHEP {\bf 0911}, 002 (2009)
  [arXiv:0907.2189 [hep-th]].




\bibitem{Gaiotto:2009we}
  D.~Gaiotto,
  ``N=2 dualities,''
  JHEP {\bf 1208}, 034 (2012)
  [arXiv:0904.2715 [hep-th]].




\bibitem{Terashima:2012ra}
  S.~Terashima,
  arXiv:1207.2163 [hep-th].


\bibitem{Witten:1997sc}
  E.~Witten,
  Nucl.\ Phys.\ B {\bf 500}, 3 (1997)
  [hep-th/9703166].


\bibitem{ArkaniHamed:2001ie}
  N.~Arkani-Hamed, A.~G.~Cohen, D.~B.~Kaplan, A.~Karch and L.~Motl,
  ``Deconstructing (2,0) and little string theories,''
  JHEP {\bf 0301}, 083 (2003)
  [hep-th/0110146].

\bibitem{Lambert:2012qy}
  N.~Lambert, C.~Papageorgakis and M.~Schmidt-Sommerfeld,
  ``Deconstructing (2,0) Proposals,''
  Phys.\ Rev.\ D {\bf 88}, 026007 (2013)
  [arXiv:1212.3337].


\bibitem{Hama:2012bg}
  N.~Hama and K.~Hosomichi,
  ``Seiberg-Witten Theories on Ellipsoids,''
  JHEP {\bf 1209}, 033 (2012)
  [Addendum-ibid.\  {\bf 1210}, 051 (2012)]
  [arXiv:1206.6359 [hep-th]].


\bibitem{Cordova:2013bea}
  C.~Cordova and D.~L.~Jafferis,
  ``Five-Dimensional Maximally Supersymmetric Yang-Mills in Supergravity Backgrounds,''
  arXiv:1305.2886 [hep-th].




\bibitem{Gaiotto:2008sa}
  D.~Gaiotto and E.~Witten,
  ``Supersymmetric Boundary Conditions in N=4 Super Yang-Mills Theory,''
  J.\ Statist.\ Phys.\  {\bf 135}, 789 (2009)
  [arXiv:0804.2902 [hep-th]].


\bibitem{Cordova:2013cea}
  C.~Cordova and D.~L.~Jafferis,
  arXiv:1305.2891 [hep-th].




\bibitem{Pestun:2009nn}
  V.~Pestun,
  JHEP {\bf 1212}, 067 (2012)
  [arXiv:0906.0638 [hep-th]].



\bibitem{Maldacena:2002rb}
  J.~M.~Maldacena, M.~M.~Sheikh-Jabbari and M.~Van Raamsdonk,
  JHEP {\bf 0301}, 038 (2003)
  [hep-th/0211139].

\bibitem{Lin:2005nh}
  H.~Lin and J.~M.~Maldacena,
  Phys.\ Rev.\ D {\bf 74}, 084014 (2006)
  [hep-th/0509235].


\bibitem{Qiu:2013aga}
  J.~Qiu and M.~Zabzine,
  Phys.\ Rev.\ D {\bf 89}, 065040 (2014)
  [arXiv:1312.3475 [hep-th]];
  J.~Qiu, L.~Tizzano, J.~Winding and M.~Zabzine,
  arXiv:1403.2945 [hep-th].


\bibitem{Nieri:2013yra}
  F.~Nieri, S.~Pasquetti and F.~Passerini,
  arXiv:1303.2626 [hep-th];
  F.~Nieri, S.~Pasquetti, F.~Passerini and A.~Torrielli,
  arXiv:1312.1294 [hep-th].



\bibitem{Hollowood:2003cv}
  T.~J.~Hollowood, A.~Iqbal and C.~Vafa,
  JHEP {\bf 0803}, 069 (2008)
  [hep-th/0310272].


\bibitem{Fujitsuka:2012wg}
  M.~Fujitsuka, M.~Honda and Y.~Yoshida,
  JHEP {\bf 1301}, 162 (2013)
  [arXiv:1209.4320 [hep-th]].



\end{thebibliography}
\end{document}